\documentclass[a4paper,useAMS,usenatbib,fleqn]{mn2e}
\usepackage{graphics,graphicx}
\usepackage{amsmath}
\usepackage{amssymb}
\usepackage{aas_macros}
\usepackage[usenames,dvipsnames]{xcolor}

\newcommand{\sigmasfr}{\dot \Sigma_{\rm \star}}
\newcommand{\csigmasfr}{\dot\Sigma_{\rm \star,0}}
\newcommand{\sigmasfrcrit}{\dot\Sigma_{\rm \star,crit}}
\newcommand{\qii}{\hbox{$Q_{\rm H II}$}}
\newcommand{\eagle}{{\sc eagle}}

\usepackage[para]{footmisc}

\title[]{Winds of change: reionization by starburst galaxies}
\author[Sharma et al.]
{Mahavir Sharma$^1$\thanks{mahavir.sharma@durham.ac.uk}, Tom Theuns$^1$, Carlos Frenk$^1$, Richard G. Bower$^1$, \newauthor Robert A. Crain$^2$, 
Matthieu Schaller$^1$ \& Joop Schaye$^3$\\
$^{1}$ Institute for Computational Cosmology, Department of Physics, University of Durham, South Road, Durham, DH1 3LE, UK\\
$^{2}$ Astrophysics Research Institute, Liverpool John Moores University, 146 Brownlow Hill, Liverpool L3 5RF, UK\\
$^{3}$ Leiden Observatory, Leiden University, P.O. Box 9513, 2300 RA Leiden, the Netherlands
}

\begin{document}

\date{Submitted ---------- ; Accepted ----------; In original form ----------}


\maketitle
\begin{abstract}
  We investigate the properties of the galaxies that
  reionized the Universe and the history of cosmic reionization using
  the ``Evolution and Assembly of GaLaxies and their Environments''
  (\eagle) cosmological hydrodynamical simulations. We obtain the
  evolution of the escape fraction of ionising photons in galaxies assuming that galactic winds create channels through
  which 20~percent of photons escape when the local surface density of
  star formation is greater than
  $0.1$~M$_\odot$~yr$^{-1}$~kpc$^{-2}$. Such threshold behaviour for
  the generation of winds is observed, and the rare local objects
  which have such high star formation surface densities exhibit high
  escape fractions of $\sim10$~percent.  
  In our model the luminosity-weighted mean
  escape fraction increases with redshift as 
  $\bar f_{\rm esc}=0.045~((1+z)/4)^{1.1}$ at $z>3$, and the galaxy number weighted mean as $\langle f_{\rm esc} \rangle=2.2\times10^{-3}~((1+z)/4)^4$, and becomes constant $\approx0.2$ at redshift $z>10$. The escape fraction evolves as an increasingly large fraction of stars forms
  above the critical surface density of star formation at earlier
  times. This evolution of the escape fraction, combined with that of
  the star formation rate density from \eagle, reproduces the inferred
  evolution of the filling factor of ionised regions during the
  reionization epoch ($6<z<8$), the evolution of the post-reionization
  ($0\leq z<6$) hydrogen photo-ionisation rate, and the optical depth
  due to Thomson scattering of the cosmic microwave background photons
  measured by the Planck satellite.
\end{abstract}
\section{Introduction}

The hot Universe emerging from the Big Bang cooled as it expanded and,
at a redshift of $z\sim 1100$, reached a temperature of $T\sim
3000$~K, cold enough for the hydrogen in the primordial gas to become
neutral \citep{Mather94}. By a redshift of $z\sim 100$, the
temperature had dropped further so that the radiation temperature
corresponded to infrared wavelengths and the Universe became
effectively dark. These \lq Dark Ages\rq\ came to an end as hot stars
in the first galaxies emitted photons with energies greater than the
ionisation potential of hydrogen, (re)ionising and heating the gas to
a temperature of $T\sim 10^4$~K. Other sources of photons could have
played a role as well\footnote{For reviews on the science of reionization see {\em
    e.g.} \cite{Barkana01,Fan06}; for recent discussions of how this science 
    drives the design of future observatories, see {\em
    e.g.} \cite{Stiavelli09} for the {\em James Webb Space Telescope}
  (JWST); and \cite{Mellema13} and \cite{,Iliev15} for the {\em Square
    Kilometer Array} (SKA).}.

Regions around the first galaxies were ionised first, growing in
number and size until they eventually coalesced as more and brighter
star-forming galaxies appeared \citep[e.g][]{Miralda00,Gnedin00,Ciardi00,Keller14, Furlanetto16}.
Therefore, as reionization
proceeded, an increasingly large fraction of the Universe became
ionised. The temporal history of reionization can therefore be
quantified by the evolution of the volume filling factor of ionised
gas as a function of redshift, \qii$(z)$.

Measurements of the column density of electrons between us and the
surface of last scattering (the Thomson optical depth,
e.g. \citealt{Planck16}), the frequency of patches containing neutral
gas as measured in quasar spectra (\lq dark gaps\rq,
\citealt{McGreer15}), the rapid evolution in the number density of
galaxies detected in Lyman-$\alpha$ emission and their clustering
\citep[e.g.][]{Caruana12,Tilvi14,Ono12,Schenker14,Ouchi10,Ota08}, the
presence of neutral gas around high redshift quasars
\citep[e.g.][]{Mortlock11} and gamma-ray bursts
\citep[e.g.][]{Totani14}, are all consistent with a Universe that is
mostly neutral ($Q_{\rm HII}\sim 0$) at $z\gtrsim 8$ and highly
ionised ($Q_{\rm HII}\sim 1$) at $z\lesssim 6$, see
\cite{Robertson15,Bouwens15}, and Fig.\ref{fig_Q} below. A more quantitative
characterisation of this ionisation phase transition may require detection
of the signal of residual neutral hydrogen (\ion{H}{i}) with the next
generation of radio telescopes.

Once a hydrogen atom has been ionised, it can recombine with a free
electron. The equation describing the evolution of $Q_{\rm HII}$
therefore contains a term describing photo-ionisation and a term
describing recombination \cite[e.g.][]{Haardt12},
\begin{equation}
\dot Q_{\rm HII}={\bar f_{\rm esc}\dot n_{\gamma,\star}\over \langle n_{\rm H}\rangle} - 1.08\,\alpha_{\rm B}\,{\cal C}\langle n_{\rm H}\rangle Q_{\rm HII}, 
\label{eq:Q2}
\end{equation}
where $\dot n_{\gamma,\star}$ is the rate at which ionising photons
are produced per unit volume (it is proportional to the cosmic star
formation rate density: $\dot
n_{\gamma,\star}=f_\star\,\dot\rho_\star$, where $f_\star$ depends on
the properties of the stellar population such as its initial mass
function (IMF)); $\bar f_{\rm esc}$ is the mean fraction of io
photons that escape from galaxies; $\langle n_{\rm H}\rangle$ is the
mean hydrogen number density; $\alpha_{\rm B}$ is the recombination
coefficient; the factor 1.08 accounts for the reionization of
\ion{He}{i} to \ion{He}{ii}; and ${\cal C}\equiv \langle n_{\rm
  H}^2\rangle/\langle n_{\rm H}\rangle^2$ is the clumping factor
\citep[e.g.][]{Pawlik09} that takes into account absorption by Lyman
limit systems (LLSs). A complete radiative transfer calculation is
required to estimate rigorously the effect of LLSs and accounting for 
spatial variations in the ionised filling factor
\citep[e.g.][]{Shukla16}. However, Eqn.~(\ref{eq:Q2}) is used
extensively and provides a reasonable description of the
global reionization history \citep{Haardt12,Haardt15,Robertson15,Bouwens15,Khaire15,Mitra15,Gnedin16}.

After reionization the Universe is highly ionised with islands of
neutral, or almost neutral gas. The distribution of these neutral
absorbers sets the mean free path of an ionising photon, with the
higher column density absorbers (neutral column $N_{\rm H{\sc
    I}}\gtrsim 10^{17}$~cm$^{-2}$) acting as photon sinks which
determine the hydrogen photo-ionisation rate, $\Gamma_{\rm HI}$, for a
given photon emissivity, $\bar f_{\rm esc}\,\dot n_\gamma$. Many
models of reionization simply tune $\bar f_{\rm esc}\dot n_\gamma$ to
reproduce the observed evolution of $\dot Q_{\rm HII}$ but a
successful model should also reproduce the observed value of
$\Gamma_{\rm HI}$ at redshifts lower than the redshift of
reionization.

The low observed value of the escape fraction in present-day galaxies
(for example, \citealt{Bland-Hawthorn01} find at most a few percent for
the Milky Way), and the measured evolution of $\dot n_\gamma$, $Q_{\rm
  HII}$, and $\Gamma_{\rm HI}$ over the redshift range
$z=3\hbox{--}8$, are inconsistent with $\bar f_{\rm esc}$ remaining
constant in time. Instead, the escape fraction should increase rapidly
with redshift, $\bar f_{\rm esc}\approx 1.8\times
10^{-4}\,(1+z)^{3.4}$, according to \cite{Haardt12} \citep[see
also][]{Khaire15,Gnedin16,Price16,Faisst16}. Consistent with this,
\cite{Bolton07} and \cite{Pawlik09} claim that the escape fraction of
galaxies at $z\sim 6$ should be of order 20~percent to account for the
(high) ionisation level of the hydrogen gas in the intergalactic
medium (IGM). Few, if any, observed values of $f_{\rm esc}$ in
galaxies at $z<3$ are as high as 20~percent.

Direct evidence that the escape fraction increases with redshift is
somewhat inconclusive; claimed detections of ionising photons leaking
from galaxies are controversial
\citep{Mostardi15}. \citet{Bridge10,Siana10} and \citet{Rutkowski15}
detect no ionising photons escaping at all from galaxies at $z\sim 1$,
with 3$\sigma$ upper limits of order of 2~percent. Typically 10\% of
the $z\sim 3$ Lyman break galaxies (LBGs) show detectable escaping
Lyman-continuum photons ($f_{\rm esc}\sim 10$~percent;
\citealt{Siana07, Siana10}). \cite{Iwata09} report similar values in a
Subaru deep field survey, as do \cite{Nestor13} from HST
data. \cite{Vanzella12} report a detection $f_{\rm esc}\approx
24$~percent from a $z=4$ LBG. \cite{Matthee16} claim detection of escape fractions as high as $60$~percent from brighter galaxies in their sample at  $z\geq2$. \cite{Leitet13} claim that observed escape fractions are higher for galaxies with higher specific star
formation rates. Observations by \cite{Zastrow11,Zastrow13} indicate
that ionising photons escape through cones presumably created by
galactic winds.

Numerical simulations that include radiative transfer face the
formidable challenge of modelling accurately the structure of the
absorbing interstellar medium \citep[e.g.][]{Pawlik15,Pawlik16} and
simulations performed by different groups yield rather
contradictory results, ranging from a very low value of a few percent
\citep[e.g.][]{Gnedin08} to very high values of $f_{\rm
  esc}=80$~percent during starbursts at high redshift
\citep[e.g.][]{Wise09,Wise14}. Furthermore \cite{Wise09} find the
escape fraction to increase with increasing halo mass but 
\cite{Razoumov10,Yajima11,Paardekooper15} find the opposite
trend. 

What these simulations do have in common is that the feedback from
supernovae (SNe) associated with recent star formation generates large
density contrasts in the interstellar medium (ISM), puncturing
channels through which winds - and presumably also ionising photons
escape
\citep[e.g.][]{Razoumov06,Gnedin08,Wise09,Yajima11,Trainor15,Ma15,Pawlik15,Pawlik16}, 
a fact that is also indicated by the dependence of escape fraction on
specific star formation rate commonly reported in such simulations. Such
outflows are observed in $z\sim 3$ LBGs \citep[e.g.][]{Pettini02},
lower redshift $z\sim 1$ galaxies \citep[e.g.][]{Erb12}, and in local
starbursts \citep[e.g.][]{Strickland09}. Simulations of galaxy
formation generate winds and these regulate star formation, 
with large mass-loading factors in lower-mass galaxies
($M_\star<10^{10}$~M$_\odot$) to avoid overproducing the faint-end of
the galaxy luminosity function \citep[e.g.][]{White78,White91}. It is
these winds that are thought to enrich the low-density intergalactic
medium (IGM) with metals \citep[e.g.][]{Madau01,Theuns02,Booth12}.

\citet[][see also \citealt{Heckman02}]{Heckman01} used H$\alpha$, NaD
and X-ray observations to claim that efficient winds are launched
provided the surface density of star formation, $\sigmasfr$, exceeds a
threshold value of $\sigmasfrcrit=0.1$~M$_\odot$~yr$^{-1}$~kpc$^{-2}$,
a finding that is also supported by theoretical studies
\citep[e.g.][]{Murray11,Scan12,Scan13}. If such winds are indeed
instrumental in clearing the path for the escape of ionising photons,
then we would expect galaxies with strong winds to have high(er)
escape fractions \citep{Heckman11} - at least in the absence of
absorption by dust, likely a good approximation at the high redshifts
($z>6$) of the reionization era. The few detections of large escape
fractions do indeed correspond to galaxies undergoing an intense
starburst with high star formation rates, $\dot M_\star \gtrsim
10$~M$_\odot$ yr$^{-1}$ occurring in a compact region (area $S\approx
(1{\rm kpc})^2$), hence $\dot M_\star/S \gg 0.1$~M$_\odot$ yr$^{-1}$
kpc$^{-2}$
\citep{Borthakur14,deBarros15,Izotov16,Izotov16b}. Heckman's model
would suggest that these galaxies should drive outflows. This is the
case for galaxy J0921+4509 discussed by \cite{Borthakur14} which
indeed drives a strong wind \citep{Heckman11}; it would be very
interesting to verify this feature in other galaxies with large
$f_{\rm esc}$.

Existing work has not yet integrated these ideas into models of
reionization. Instead, analytic or semi-analytic studies generally assume a
constant value of $f_{\rm esc}$ for all galaxies at a given redshift
\citep[e.g.][]{Robertson15,Bouwens15}. We have developed a formalism, 
described in 
Sections 3 and 4, in which the escape fraction of
ionising photons from a galaxy is closely intertwined with its
feedback activity and therefore varies from galaxy to galaxy. We first
outlined and applied this formalism in \cite{Sharma15} to identify the
galaxies that contribute the most to the ionising radiation and found 
that the brighter galaxies at high redshift, which are above the HST
detection limit, provide a large fraction ($\approx 50$ percent) of
the ionising photons responsible for reionization.

In this paper we investigate the history of reionization analysing
galaxies from the ``Evolution and Assembly of GaLaxies and their
Environments'' (\eagle) suite of cosmological hydrodynamical
simulations \citep{Schaye15, Crain15}. We infer the evolution of the
escape fraction, and this enables us to calculate the evolution of the
filling factor of ionized gas and the electron scattering optical
depth and, crucially, verify that we obtain a realistic evolution for
the post-reionization amplitude of the UV-background. The results are
particularly interesting in the aftermath of the recently revised
lower values of the optical depth by \cite{Planck16},
implying that the Universe was reionized rather later than previously
thought \citep[e.g.][]{Haardt12}.

This paper is organised as follows. In Section 2 we describe the
implementation of subgrid physics in the \eagle\ simulation,
particularly the aspects that are relevant to this paper. In Section 3
we compare the star-formation history from \eagle\ to the
observations, and study the evolution of the star formation rate and
related quantities. We then explore the properties of the star-forming
regions within galaxies, and derive the escape fraction of ionising
photons in Section 4.  In Section 5 we compute the ionising
emissivity, the photo-ionisation rate and the electron scattering
optical depth. We compare our results with observations and discuss
our findings in the final section.

\section{The {\sc eagle} simulations}
\label{sec_eagle}
In this work we use the \eagle\ suite of cosmological hydrodynamical
simulations \citep{Schaye15, Crain15}. These were
performed with the {\sc gadget-3} implementation (last described by
\citealt{Springel05}) of the tree-SPH algorithm, with \lq subgrid\rq\
modules for physical processes below the resolution limit. We briefly
review these modules here, paying particular attention to the star
formation and reionization implementations, since these are the most
relevant to this paper.

The simulations are performed in cubic periodic volumes and start
redshift of $127$ from cosmological initial conditions generated using
second order Lagrangian perturbation theory, as described by
\cite{Jenkins13a}. We use the \cite{Planck14} values of the
cosmological parameters. The simulation takes advantage of
improvements to the basic SPH and timestepping algorithms implemented
in {\sc gadget-3}, to which we collectively refer as {\sc
  anarchy}. These are described by \cite{Schaye15}, with
\cite{Schaller15} illustrating their relatively small effects on the
properties of galaxies. The simulation tracks 11 elements (H, He, C,
N, O, Ne, Mg, Si, S, Ca, Fe) released during the evolution of massive
stars, asymptotic giant branch (AGB) stars and type~I and type~II
supernovae (SNe), as described in \cite{Wiersma09a}.

Element-by-element cooling and photo-heating by the optically thin
UV/X-ray background of \cite{Haardt01}, including Compton cooling
and thermal Bremsstrahlung,
are implemented as described by \cite{Wiersma09b}. We assume here that
hydrogen reionizes at redshift\footnote{The epoch of reionization imposed on the simulation is therefore not consistent with what we estimate is the actual epoch of reionization. We verified that this does not impact our conclusions significantly: in these simulations, the effects of reionization on the cosmic star formation rate is relatively small.} $z=11.5$, when we inject an extra 2~eV
of energy per hydrogen atom to take account of non-equilibrium effects
that occur when neutral gas is overrun by an ionisation front
\citep{Abel99}. At higher $z$ we impose a uniform radiation
field but ignore photons above 1~Rydberg; this prevents the formation
of molecules that are neither tracked nor resolved but would affect
the cooling rate.

Star formation is implemented by stochastically converting gas
particles into collisionless star particles, at a rate set by the
local gas pressure \citep{Schaye08}. This reproduces the observed
$z=0$ Kennicutt-Schmidt relation \citep{Kennicutt98}, and we assume
that this relation does not evolve. The star formation rate is assumed
to be zero below the metallicity-dependent threshold of
\cite{Schaye04} for the formation of a cold ($T\ll10^4$ K) gas
phase. Feedback from star formation is implemented thermally as
described by \cite{DallaVecchia12}. The simulation also models the
formation, growth, and feedback from supermassive black holes, as
described by \cite{Springel05b,Booth09}, but taking into account the
angular momentum of accreting gas, as described by
\cite{RosasGuevara15}.

The free parameters of the subgrid modules are calibrated using the
observed $z\approx 0.1$ galaxy stellar mass function, galaxy sizes,
and the relation between black hole mass and stellar mass, as
described by \cite{Crain15}. \eagle\ reproduces a wide range of
observables that were not part of the calibration process. Most
relevant to this work are the papers by \cite{Furlong15a} and
\cite{Furlong15b} that show the \eagle\ broadly reproduces the
evolution of galaxy masses and sizes, respectively.

{ In this paper we use three simulations from the \eagle\ suite that
differ in volume to verify that our results are converged, L0025N0376,
L0050N0752, and L0100N1504 in Table~2 in \cite{Schaye15}. These adopt
identical subgrid models and parameters and are performed
in volumes of 25, 50 and 100 co-moving megaparsecs on a side,
respectively. Gas particles in simulations L025N0376 and L0100N1504 have
initial particle masses of $m_g=1.81\times 10^6$ M$_\odot$. The co-moving Plummer equivalent
gravitational softening length is $\epsilon_{\rm com}=2.66$~kpc and the maximum physical 
gravitational softening length is $\epsilon_{\rm prop}=0.70$~kpc. In the higher resolution 
simulation, L0050N0752, $m_g=2.26\times 10^5$ M$_\odot$, $\epsilon_{\rm com}=1.33$~kpc and
$\epsilon_{\rm prop}=0.35$kpc.}

\section{The evolution of the star formation surface density in galaxies}
\label{prelude}
\begin{figure}
 \centering
 \includegraphics[width=\linewidth]{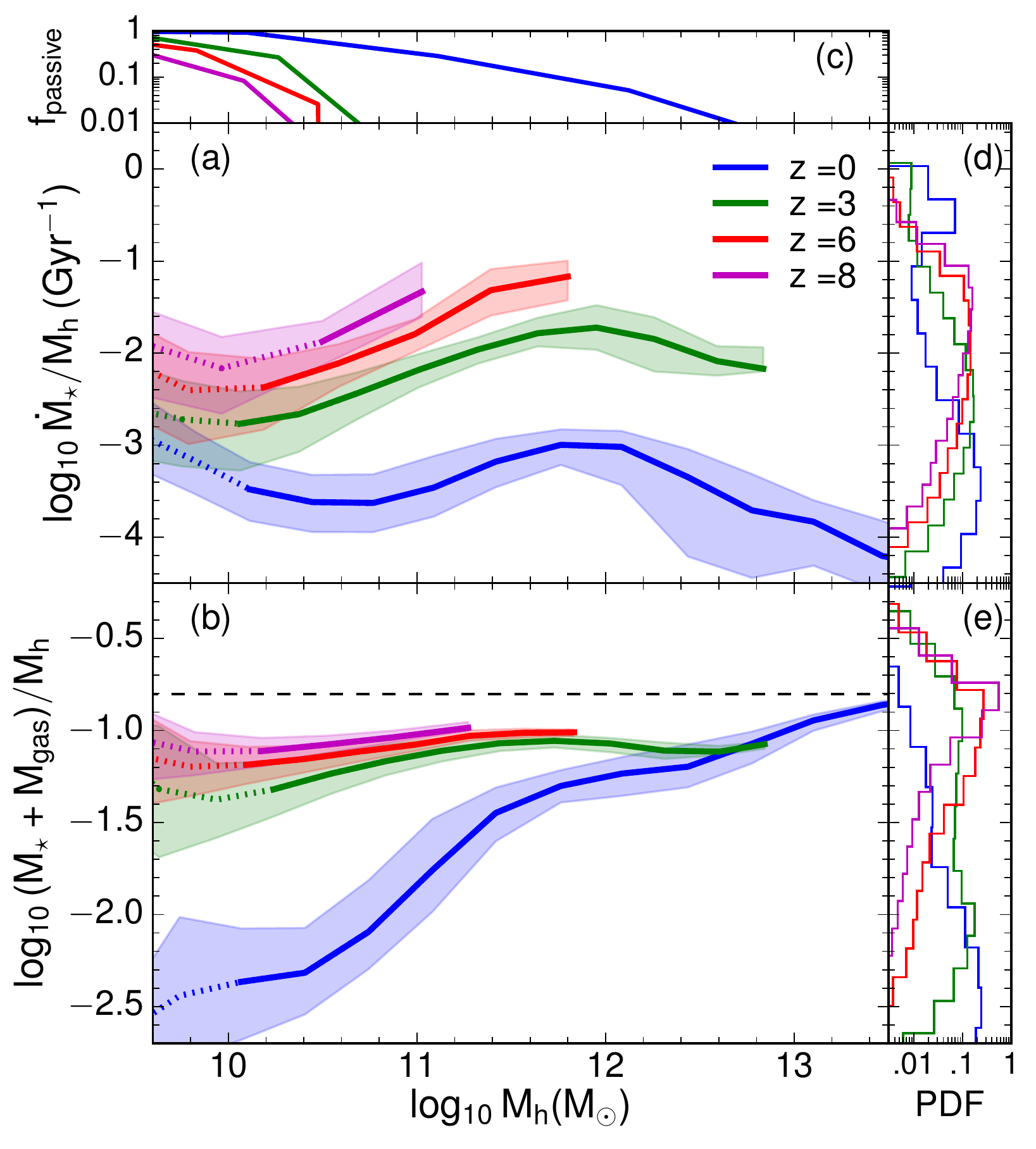}
 \caption{The median star formation rate ($\dot M_\star$; {panel a})
   and the halo baryon fraction ({panel b}) as a function of the halo
   mass ($M_{\rm h}$) for galaxies in the \eagle\ simulation, shown
   as solid curves for redshift 0 (blue), 3 (green), 6 (red) and 8
   (magenta). Shaded regions show the range of 25th to 75th
   percentile. The corresponding passive fractions are shown in panel
   c. The probability distribution of $\dot M_\star/M_{\rm h}$ is
   plotted in panel d and that of the baryon fraction in panel e. The
   halo baryon fraction and star formation rate both increase rapidly
   with redshift.}
 \label{fig_hist1}
\end{figure}

\begin{figure}
 \centering
 \includegraphics[width=\linewidth]{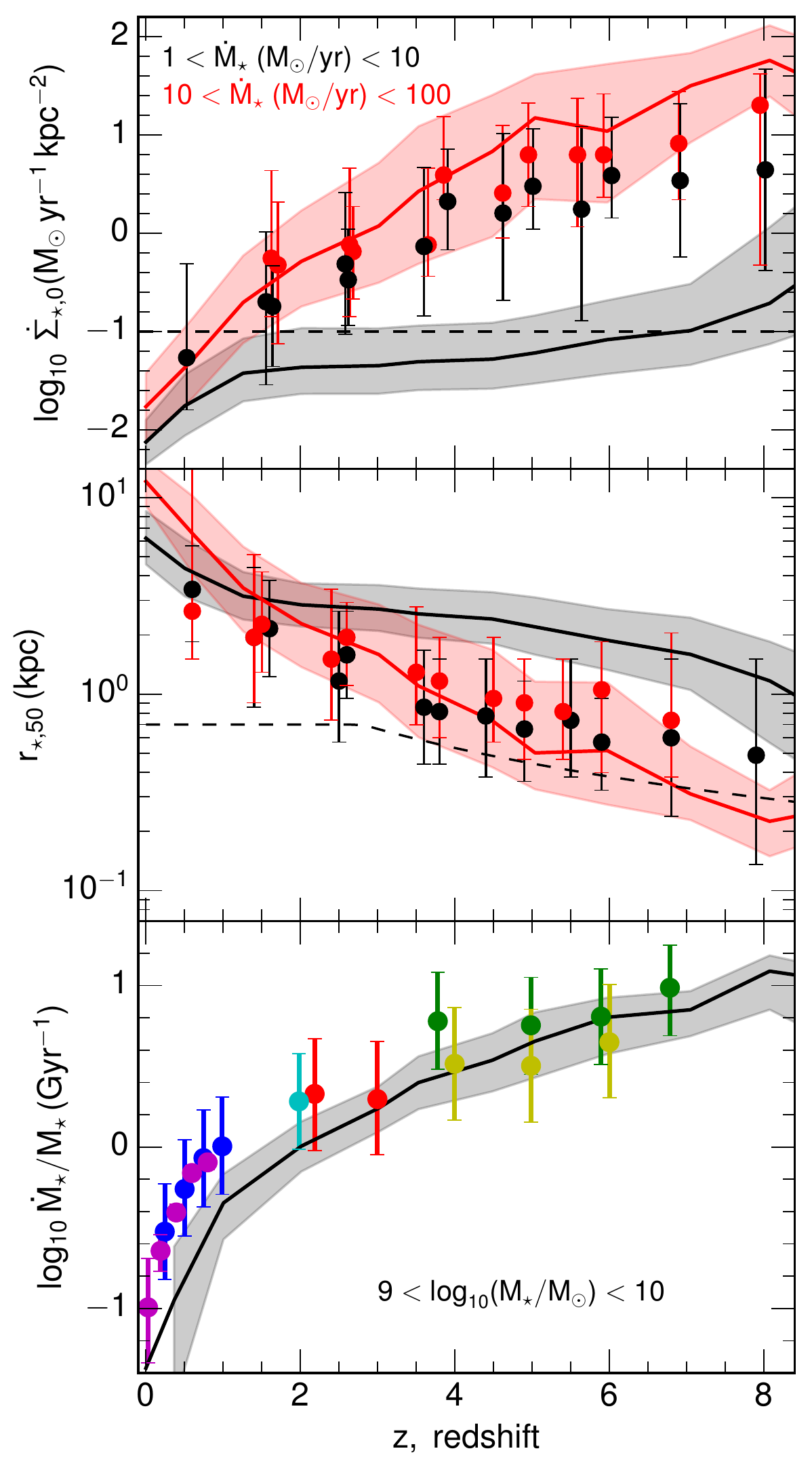}
 \caption{Evolution of the central star formation rate surface
   density, $\csigmasfr$ ({\em top panel}), and of the stellar
   half-mass radius, $r_{\star, 50}$ ({\em middle panel}) for \eagle\
   galaxies with star formation rate $1<\dot M_\star/{\rm
     M}_\odot\,{\rm yr}^{-1}<10$ (black) and $10<\dot M_\star/{\rm
     M}_\odot\,{\rm yr}^{-1}<100$ (red); solid lines show the median
   relation with the shaded area including the 25-75th
   percentiles. Symbols with $1~\sigma$ error bars are observations 
   from \citet{Shibuya15}, with colours corresponding to the same
   limits in $\dot M_\star$. The horizontal dashed line in the top
   panel denotes the threshold for driving outflows \citep{Heckman01}
   and the dashed line in the middle panel shows the gravitational
   softening length. There is general agreement between \eagle\ and
   the measurements in that $\csigmasfr$ increases rapidly with $z$,
   although \eagle\ galaxies with lower star formation rate tend to be
   larger and have lower values of $\csigmasfr$ than observed at
   $z>2$.  The bottom panel compares the evolution of $\dot
   M_\star/M_\star$ for \eagle\ galaxies with $10^9<M_\star/{\rm
     M}_\odot<10^{10}$ (black line: median relation, shaded area
   25-75th percentiles) with observational data from
   \protect\cite{Noeske07} (blue), \protect\cite{Damen09} (magenta),
   \protect\cite{Reddy09} (red), \protect\cite{Stark13} (green),
   \protect\cite{Gonzalez14} and \protect\cite{Daddi07} (cyan). The
   specific star formation rate in \eagle\ increases rapidly with $z$,
   tracking the observations, but is low by a factor of $\sim
   0.2-0.3$~dex below $z=1$. }

\label{fig_sigsize}
\end{figure} 
In this section we demonstrate that the surface density of star
formation averaged on kiloparsec scales, $\sigmasfr$, increases
rapidly with redshift in \eagle, examine whether there is
observational support for this trend, and investigate the underlying
physical processes that drive the evolution in the simulation.

The star formation rate of star-forming galaxies as a function of halo
mass\footnote{By halo mass, we mean the mass enclosed by a sphere, within which the mean density is 200 times the critical density.}, $M_{\rm h}$, and the baryon fraction, $f_{\rm baryon}\equiv
(M_\star+M_{\rm gas})/M_{\rm h}$, of \eagle\ galaxies are plotted at
several redshifts in Fig.\ref{fig_hist1}. At $z=0$, $f_{\rm baryon}$
increases strongly with $M_{\rm h}$ up to $M_{\rm h}\approx
10^{12}{\rm M}_\odot$, and continues to increase above that but at
a slower rate. The corresponding star formation rate $\dot
M_\star/M_{\rm h}$ remains approximately constant up to $M_{\rm
  h}=10^{12}~{\rm M_\odot}$ and declines at higher $M_{\rm h}$.  Clearly, $\dot
M_\star/M_{\rm h}$ and $f_{\rm baryon}$ do not track each other well
because star formation is self-regulating with the dense gas fraction
set by the efficiency of feedback, as demonstrated by \cite{Haas13}
for galaxies and by \cite{Altay13} for damped Lyman-$\alpha$ systems
in the {\sc owls} simulations presented by \cite{Schaye10}. In
particular, the decline in $\dot M_\star/M_{\rm h}$ above $M_{\rm
  h}=10^{12}{\rm M}_\odot$ is due to AGN feedback in \eagle. The
feedback in \eagle\ is calibrated to reproduce the $z=0$ stellar mass
function and the sizes of galaxies. The simulation then also
reproduces the star formation rate as a function of $M_\star$, with star
forming galaxies having the right colours and with approximately the right
fraction of passive red galaxies \citep{Trayford15,Trayford16}.
{ Even though \eagle\ reproduces well the observed increase in
specific star formation rate (sSFR), $\dot M_\star/M_\star$, with $z$,
the values of the sSFR as a function of $M_\star$ are typically too low by a factor of two, 
as is the star formation rate density (\citet{Furlong15a}, their Figs.~4 and 5).
\cite{Furlong15a} show that, nevertheless, \eagle\ reproduces the observationally inferred stellar mass functions
well. This may appear contradictory, as the integral of the star formation rate gives
the stellar mass. \cite{Mitchell13} examined the origin of this apparent discrepancy
in the semi-analytical {\sc galform} model, concluding that several observational
biases may play a role.}

Towards higher $z$ both $f_{\rm baryon}$ and $\dot M_\star/M_{\rm h}$
increase rapidly in \eagle, while the passive fraction decreases. This leads to
the blue cloud of star forming galaxies becoming bluer with increasing
$z$, and the red sequence thinning out, as shown by \cite{Trayford16}
up to $z=2$. This trend continues to higher $z$, with for example the
star formation rate of a galaxy hosted by a halo of mass $M_{\rm
  h}=10^{11}{\rm M}_\odot$ increasing by nearly two orders of
magnitude from $z=0$ to $z=6$. These higher star formation rates with
increasing $z$ result in a luminosity function of \eagle\ galaxies
that is in good agreement with observations out to $z=7$, as shown by
\cite{Furlong15a}, and potentially even higher $z$, as shown in
Fig.~\ref{fig_sigsize}.

The strong evolution of the star formation rate at fixed halo mass is
a consequence of a balance between the much higher rate of
cosmological accretion at higher $z$, and the evolution of the
efficiency of feedback from star formation (and black holes at higher
$M_{\rm h}$). In particular, feedback needs to be {\em more} efficient
at higher $z$ to avoid producing too many stars early on. Conversely,
we find that feedback needs to be relatively inefficient at low $z$,
because otherwise the specific star formation rate would be lower than
observed. In {\sc eagle} these trends emerge through the dependence of
the subgrid feedback efficiency on the local gas density and
metallicity, see \cite{Crain15} for an in-depth discussion.

While the specific star formation rates of galaxies increase with $z$,
their sizes decrease (Fig.~\ref{fig_sigsize}): for a galaxy with
$10^9<M_\star/{\rm M}_\odot<10^{10}$, the specific star formation rate
$\dot M_\star/M_\star$ increases by two orders of magnitude from $z=0$
to $z=8$, while its stellar half-mass radius $r_{\star, 50}$ {\em
  decreases} by a factor of $\approx 5$ following approximately the
scaling $r_{\star, 50}\propto H(z)^{-2/3}$, where $H(z)$ is the Hubble
constant. The evolutionary trends in $\dot M_\star/M_\star(z)$ and
for $r_{\star, 50}(z)$, seen in the simulation, are observed as well. \cite{Kawata15} infer
half-mass radii of $\approx$~kiloparsec size for bright $z=6-8$
galaxies, somewhat smaller than those of the more massive \eagle\
galaxies, with a slightly steeper redshift dependence of $\propto
(1+z)^{-1.24\pm 0.1}$ over the redshift range $z=2.5\hbox{--}12$.

With the star formation rate in galaxies increasing with $z$ and their sizes decreasing, the surface density of star formation, $\sigmasfr$, increases rapidly, by two orders of magnitude for a galaxy with $\dot M_\star\sim 10~{\rm M}_\odot$ yr$^{-1}$ between $z=0$ and $z=6$ (Fig.~\ref{fig_sigsize}). If a high value of $\sigmasfr$ is required to drive strong winds, then we would expect that the fraction of galaxies that drive winds increases rapidly with $z$. Observations of Lyman-break galaxies indeed show that strong outflows are ubiquitous at high redshift \citep[e.g.][]{Pettini01, Shapley03, Weiner09}.

\begin{figure}
 \centering
 \includegraphics[width=\linewidth]{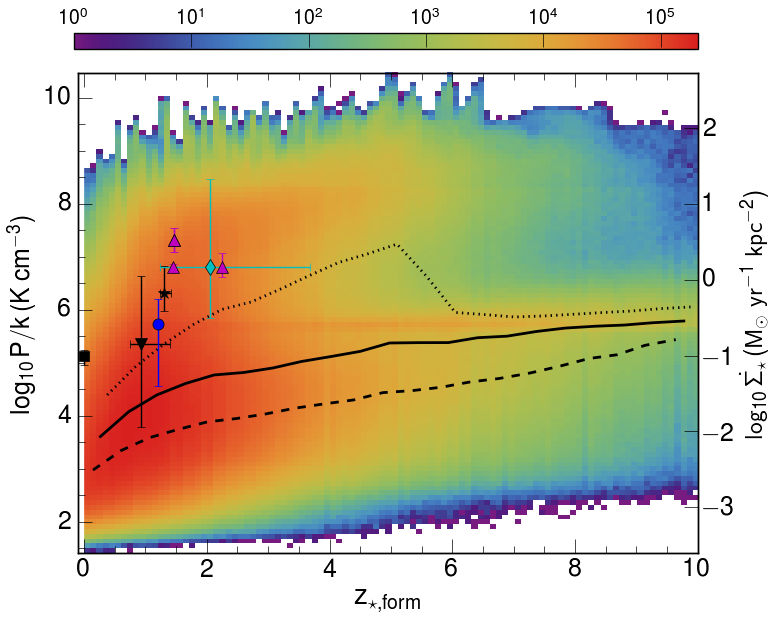}
 \caption{Probability distribution of the pressure at which stars form
   as a function of redshift. The colour coding is a measure of the
   fraction of stars that form at a given $z$ with a given pressure;
   the black solid curve shows the evolution of the median of the birth pressure. 
   The axis on the right shows the corresponding values
   of $\dot \Sigma_{\star}$ for the star particles from
   Eq.~(\ref{eq:sigmasfr}). The observations with error bars are shown as black square \citep{Kennicut03}, blue circle \citep{Freundlich13} and magenta triangles \citep{Swinbank12}. The median birth pressure increases by
   more than two orders of magnitude between $z\approx 0$ and
   $z\approx 6$, but a small fraction of stars continues to form at
   high pressure even at $z=0$. A second sequence of higher birth
   pressures appears below $z\approx 3$ due to the effects of metal
   cooling.}
 \label{fig_nPsig}
\end{figure}

What drives this evolution to higher $\sigmasfr$ at higher $z$ in \eagle? The star formation rate volume density, $\dot\rho_\star$, and surface density, $\sigmasfr$, are computed using the model of \cite{Schaye08}\footnote{The proportionality constant of Eq.~(\ref{eq:sigmasfr}) used in 
	\cite{Sharma15} is $\pi^{1/2}$ larger, because we used a different expression for the Jeans length in that paper. We verified that the lower value used here does not change any of the results
	of \cite{Sharma15}.}
,
\begin{eqnarray}
	\dot\rho_\star &\propto&  p^{(2+\gamma\,(n-1))/(2\gamma)}\approx p^{0.95}\approx \rho^{1.27}\,,
	\label{eq:rs}
	\\
	\sigmasfr &=& 0.1\,{\rm M}_\odot~{\rm yr}^{-1}\,{\rm kpc}^{-2}\,\left({p/{\rm k_B}\over 1.35\times 10^5\,{\rm K}~{\rm cm}^{-3}}\right)^{0.7}\,
	\label{eq:sigmasfr}
\end{eqnarray} where the gas pressure, $p$, is related to the gas density by the
relation, $p\propto\rho^\gamma$, and ${\rm k_B}$ is Boltzmann's
constant. The proportionality constants in Eq.~(\ref{eq:rs}) and the
exponent $n$ are derived from the Kennicutt-Schmidt law
\citep{Kennicutt98} scaled according to the Chabrier IMF used in {\sc
  eagle}. In \eagle\ we adopt $\gamma=4/3$ and $n=1.4$, which yields
the numerical values above. The evolution of $\dot M_\star$ is then a
consequence of the higher values of the pressure at which stars form,
increasing by more than two orders of magnitude between $z=0$ and
$z=6$, as shown in Fig.~\ref{fig_nPsig}. The corresponding increase in
$\sigmasfr$ is a factor of 25. A second sequence of higher pressure appears at redshifts $z<3$ that is due to the effect of metal cooling (see \citealt{Crain15} for further details).

The increase in the pressure (or equivalently density since
$\rho\propto p^{3/4}$ for star forming gas in \eagle) at which stars
form is not unique to the \eagle\ simulation. We find that in \eagle\
the median ISM density at which stars form increases by approximately
a factor of 10 between $z=0$ and $z=3$ (corresponding to an increase
in $p$ by a factor of 6); \cite{Shirazi14} find an increase by a
factor of 7$^{+10.2}_{-5.4}$ in electron density in the ISM of
galaxies in their sample between redshift $0$ and $3$, in reasonable
agreement with our results.

The observations by \cite{Genzel11} and \cite{Swinbank12} of star
forming clumps in highly star-forming galaxies at $z=1-2$ yield values
of $\sigmasfr=1-100~{\rm M}_\odot\,{\rm yr}^{-1}\,{\rm kpc}^{-2}$,
similar again to what we find in \eagle. These clumps drive strong
outflows, which, according to \cite{Genzel11}, are driven by energy
injection from massive stars and supernovae. This good agreement
between \eagle\ and these observations confirms that the phenomenology
of the physics assumed in the simulations yields realistic results,
and gives us confidence that we can reasonably use the results of the simulations to even higher redshifts where the observations are not that constraining.

We conclude that in the \eagle\ simulation stars form at increasingly
higher gas pressure with increasing $z$ and, as a consequence, the
surface density of star formation also increases rapidly. This results
in higher $z$ galaxies having higher specific star formation rates and
smaller sizes. These trends quantitatively reproduce 
observations. In real galaxies (and also in the simulation), such
regions of intense star formation drive strong winds and, in the few
cases where a measurement of the escape fraction has been possible
\citep{Borthakur14,deBarros15,Izotov16}, give rise to relatively high escape
fractions, $10\hbox{--}20$~percent. It then follows that the escape
fraction of ionising photons will tend to increase with redshift, as
an increasing fraction of stars is born in galaxies that drive strong
outflows. We examine the corresponding evolution of the escape fraction in the next section.

\section{The escape fraction of ionising photons from galaxies}
\label{sec_ParticleSFR}
\label{sec_model}
Massive stars emit ionising photons. These can be prevented
from escaping their galaxy by absorption by dust, or by being
converted to lower energies after first ionising neutral hydrogen in
the galaxy. Their escape fraction, $f_{\rm esc}$, is given by the
ratio, $\dot N_{\gamma, {\rm gal}}/\dot N_{\gamma,\star}$ of ionising
photons emitted by the {\em galaxy}, $\dot N_{\gamma, {\rm gal}}$,
over that emitted by its {\em stars}, $\dot
N_{\gamma,\star}$. Evaluating $f_{\rm esc}$ 
requires defining where the galaxy \lq ends\rq. Operationally we may
take this to mean that the photon escapes out to at least the virial
radius, $R_{200}$, of the galaxy's dark matter halo. Given the two
types of photon sinks (dust and \ion{H}{I}), we write $f_{\rm
  esc}=f_{\rm esc, dust}\times f_{\rm esc, H{\sc I}}$, but, since we
are mostly interested in the $z>6$ Universe, we take $f_{\rm esc,
  dust}=1$ \cite[see e.g.][]{Bouwens15}.

The physical reasoning that underlies our model is that strong winds
driven by massive stars carve channels through the surrounding (mostly
neutral) gas through which ionising photons can escape. There is some
evidence in very high resolution simulations that winds can indeed
result in high values of $f_{\rm esc}$, especially if star formation
is bursty \citep[e.g.][]{Wise09,Ma15}, although there is currently no
consensus on how, or even if, $f_{\rm esc}$ depends on stellar mass,
star formation rate, or redshift (compare, for example, \cite{Kimm14}
with \citealt{Paardekooper15} or \citealt{Xu16}).

Theoretically, energy injection by supernovae (SNe) creates a
thermalised hot cavity that lies at the centre of an expanding
bubble. Following \cite{Chevalier74} we find that
$\sigmasfr>10^{-1.5}~{\rm M}_\odot~{\rm yr}^{-1}~{\rm kpc}^{-2}$ is
required to achieve a filling factor close to 100~percent for the hot
media. \cite{Clark02} also find such a threshold behaviour for the
percolation of SN-blown bubbles in star forming discs. Similar limits
have been inferred from the simulations of
\cite{Fujita03,vonGlasgow13} and \cite{Scan12}. A threshold is also
a requirement for blowing radiatively driven winds from
massive star clusters \citep{Murray11} which may also play a role in
evacuating high-density gas from the surroundings of the sources of
ionising photons, leading to higher escape fractions.

We will therefore assume that the value of $f_{\rm esc}$ for individual
star forming patches in a galaxy depends on the local surface density
of star formation, $\sigmasfr$ (averaged on scales of $\approx {1~\rm
  kpc}$). As in \cite{Sharma15}, we assume the escape fraction to be
zero when $\sigmasfr<\sigmasfrcrit$, and $f_{\rm esc}=f_{\rm esc,
  max}$ when $\sigmasfr\ge \sigmasfrcrit$. We set
$\sigmasfrcrit=0.1~{\rm M}_\odot~{\rm yr}^{-1}~{\rm kpc}^{-2}$, the
critical surface density above which star forming regions are observed
to drive strong winds according to \cite{Heckman01}, and use a
default value $f_{\rm esc, max}=20$~percent, motivated by the escape
fractions observed in $z\sim 0$ starbursts by \cite{Borthakur14} and
\cite{Izotov16}. The values of $\sigmasfrcrit$ and $f_{\rm esc, max}$
are the main parameters in our model, we illustrate
how our results change if we vary them below.

Some recent studies point out that there may be a timing mismatch
between the maximum of the ionising emissivity and the peak of the
escape fraction. \cite{Kimm14} make the point that it will take some
time for massive stars to carve channels (see also \citealt{Ma16}),
and so the escape fraction may increase from very low values when the
stars form, to high values later on. However, since ionising photons
are produced throughout the life of the massive star, and the channels
through which photons escape take some time to open, only a small
fraction of all photons produced may effectively escape - even if the
{\em escape fraction} eventually reaches high values. However, this
problem may not arise for multiple bursts occurring in the same
region \citep[e.g.][]{Gentry16}. Moreover, \cite{Stanway16} argue that stellar population
synthesis models that include binary stars may play an important role,
because these binary stars have high luminosities in ionising photons
even at a time of order 100~Myr after the starburst (and after the
massive stars have been able to open up channels through which photons
can escape; see also \citealt{Ma16}). Binary stars may therefore play
an important role in setting the net emissivity of a star-bursting
galaxy.

We implement the model for the escape of photons from \eagle\ galaxies
through winds as follows. We begin by identifying young star particles
(age $<100$~Myr), and calculate the surface density of star formation
at the time they formed, $\sigmasfr$, from their birth density (using
Eq.(\ref{eq:sigmasfr}) and the pressure-density relation, $p\propto
\rho^{4/3}$, imposed on star forming gas in \eagle), which is recorded
for every star particle formed. If $\sigmasfr <\sigmasfrcrit$ we set
$f_{\rm esc}=0$ for this star-forming region, and if $\sigmasfr
\ge\sigmasfrcrit$ we set $f_{\rm esc}=f_{\rm esc,
  max}=20$~percent. Weighting each star particle by its star formation
rate at birth, we calculate the ionising luminosity-weighted value of
the escape fraction, as well as the ionising emissivity, for all \eagle\ galaxies.

\begin{figure}
 \centering
 \includegraphics[width=\linewidth]{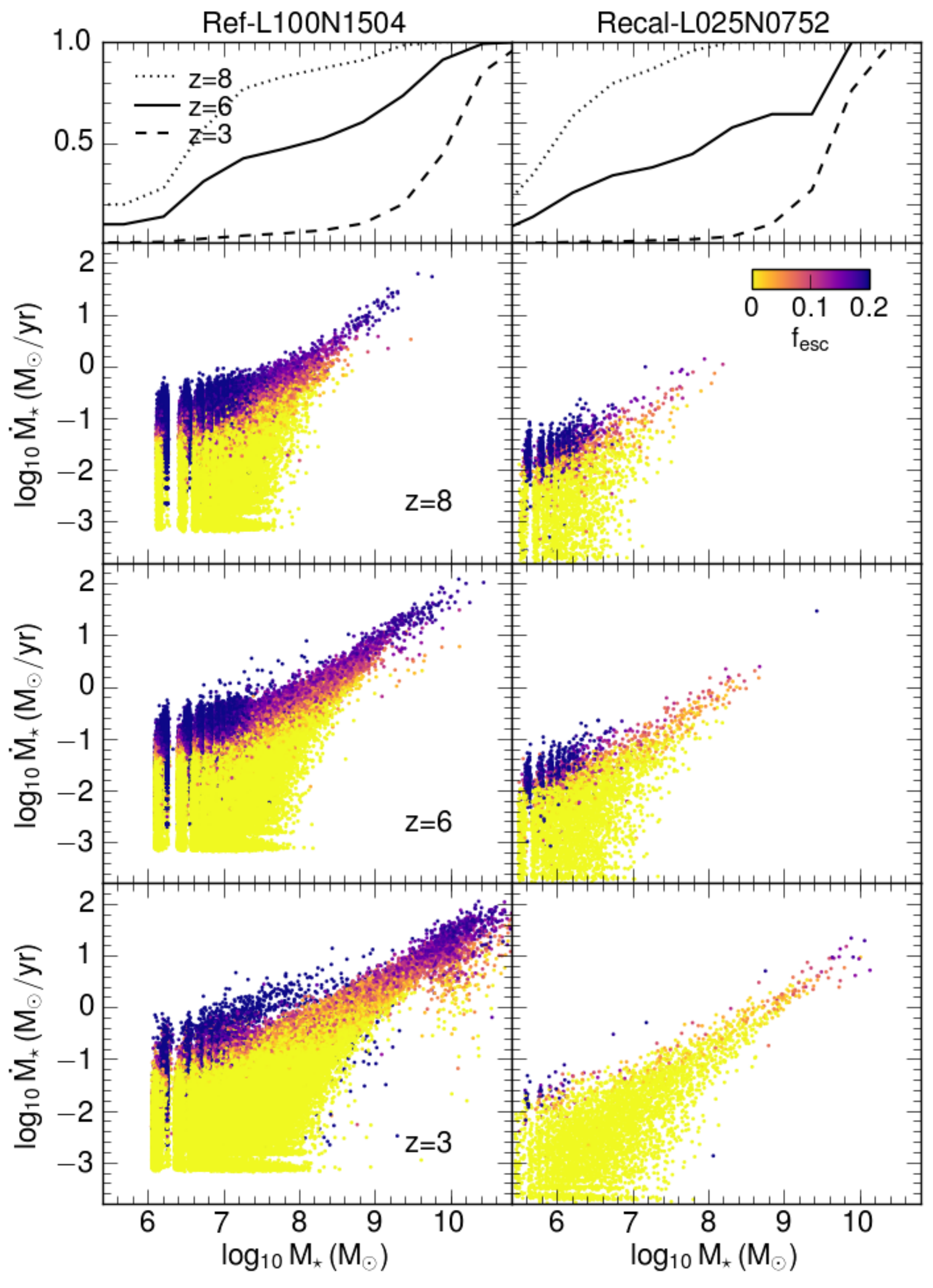}
 \caption{The escape fraction of ionising photons is shown as a function of the stellar mass ($M_\star$) and the star formation rate ($\dot M_\star$) of the galaxies in the simulation Ref-L100N1504 (left column) and Recal-L025N0752 (right column). In the top row of plots, we show the cumulative number of photons emitted by the galaxies below a given stellar mass, at redshift 3 (dashed), 6 (solid) and 8 (dotted). There is a large amount of scatter in $\dot M_\star$ at low $M_\star$, which implies that low mass does not necessarily mean faint. At a given stellar mass, the galaxies with higher star formation rate (i.e. brighter) have higher escape fractions. the top panel shows that the brighter high-mass galaxies at a given redshift dominate the emissivity. However, the contribution from  low mass galaxies increases with redshift, and at $z = 8$, approximately 50 percent of the ionising photons are emitted by galaxies below a stellar mass of $10^7$ M$_\odot$. Most of the photons emitted by these low mass galaxies arise from the brightest ones.
 }
\label{fig_fesc_mult} 
\end{figure}

{ Fig.~\ref{fig_fesc_mult} is a scatter plot of stellar mass, $M_\star$, versus star formation rate, $\dot M_\star$, for \eagle\ galaxies (points) at various redshifts, colour coded according to the instantaneous value of $f_{\rm esc}$. As expected, more massive galaxies appear at lower redshift. The left and right columns allow us to investigate the impact of resolution as well as simulation volume, since Recal-L025N0752 (right column) has eight times better mass resolution, but 64 times smaller volume than Ref-L100N1504. In both simulations, star formation is very bursty, with galaxies exhibiting a large scatter in $\dot M_\star$ at given $M_\star$.

At any value of $M_\star$, the escape fraction is highest for the most star-forming galaxies of that mass. Since such bursty galaxies are rare, they are less well sampled in Recal-L025N0752, as are the more massive galaxies. The top panels show the cumulative contribution to the total emissivity, $f(M<M_\star)$, at three redshifts: 50~per cent of the ionising emissivity comes from galaxies of mass $M_\star\lessapprox 5\times 10^6{\rm M}_\odot$, $10^8{\rm M}_\odot$ and $10^{10}{\rm M}_\odot$ at redshifts $z=8$, 6 and 3, respectively (with comparable values of $10^6$, $10^8$ and $5\times 10^9{\rm M}_\odot$ in Recal-L025N0752).

Therefore, although the stellar masses of the galaxies that produce the bulk of the ionising photons may be relatively low, their star formation rates (and hence their luminosities, since at these high redshifts galaxies are typically detected at UV rest wavelengths only) are not. At $z=8$, 50~per cent of the total ionising emissivity comes from galaxies with $M_\star\lessapprox 10^{6.5}{\rm M}_\odot$, but these have $\dot M_\star\sim 10^{-1}{\rm M}_\odot~{\rm yr}^{-1}$ and hence $M_{1500}\sim -16.5$, only just below the current HST detection limit \citep{Sharma15}. At lower $z$, brighter galaxies dominate the emissivity even more.

}

\begin{figure}
 \centering
 \includegraphics[width=\linewidth]{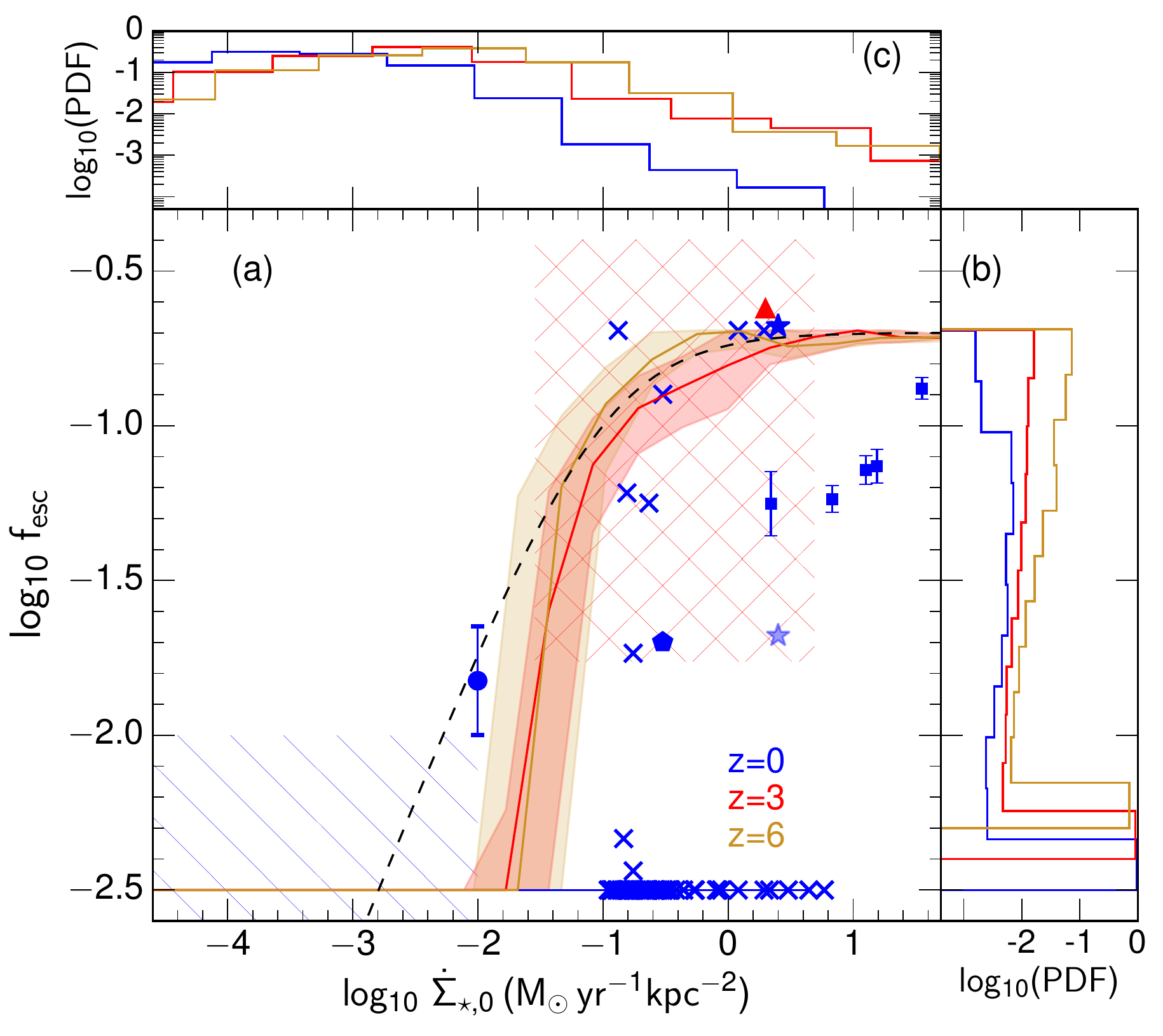}
 \caption{The galaxy-averaged escape fraction of ionising photons,
   $f_{\rm esc}$, as a function of the central surface density of star
   formation, $\csigmasfr$, at redshifts $z=0$ (blue), 3
   (red) and 6 (orange), for galaxies with
   $M_\star>10^{7}$ M$_\odot$.  The galaxies for which $f_{\rm esc}=0$ have been assigned arbitrarily a value of $\log_{10}f_{\rm esc}=-2.5$. The solid lines correspond to the median escape fractions at each redshift, with the shaded region including the 25th and 75th percentiles; the black dashed line is a simple fit to trend, and is used in the text. 
      The galaxies at $z=0$ that have $\csigmasfr>0.1~{\rm M}_\odot~{\rm yr}^{-1}~{\rm
     kpc}^{-2}$ are shown as blue crosses.  $f_{\rm esc}$ for
   galaxies with $\csigmasfr<10^{-2.5}{\rm M}_\odot~{\rm yr}^{-1}~{\rm
     kpc}^{-2}$ is typically very low or zero. For those with
   $\csigmasfr>10^{-1}{\rm M}_\odot~{\rm yr}^{-1}~{\rm kpc}^{-2}$, it
   is typically close to the maximum of 20~percent allowed by the
   model, whereas at intermediate values of $\csigmasfr$ the scatter
   in $f_{\rm esc}$ is large.  The red cross-hashed region represents
   the observed range of $z\approx 3$ Lyman-break galaxies with
   confirmed detections of ionising radiation \citep{Shapley06,Iwata09,Nestor13}, for example a recently claimed detection by \citet{Vanzella12} also lies in this region. The blue hashed region in the
   lower left corner shows the range for observed 
   galaxies at $z\approx 0$ which have low values of $f_{\rm esc} \sim 1\%$
   \citep[e.g.][]{Gnedin08}. The blue circle with error bar represents the Milky-Way \citep{Bland-Hawthorn01}, the blue pentagon corresponds to the upper limit in nearby galaxy Haro11 \citep{Grimes07}. Most $z\approx 0$
   \eagle\ galaxies have low values of $f_{\rm esc}$ although unusual
   outliers exist with high values (blue crosses), comparable to that of the galaxies found by \citet{Borthakur14} (blue star) and \citet{Izotov16,Izotov16b} (blue squares with error bars). Many of the $z\approx 3$ galaxies
   fall in the region of observed Lyman-break galaxies. Finally, a larger number of  $z=6$ galaxies have relatively high values for $f_{\rm esc}$. The
   probability distribution for $f_{\rm esc}$ and $\csigmasfr$ is shown in the  panels b and c respectively, with colours corresponding to the redshift ranges in panel a. }
\label{fig_fesc_1} 
\end{figure}
\begin{figure}
 \centering
 \includegraphics[width=\linewidth]{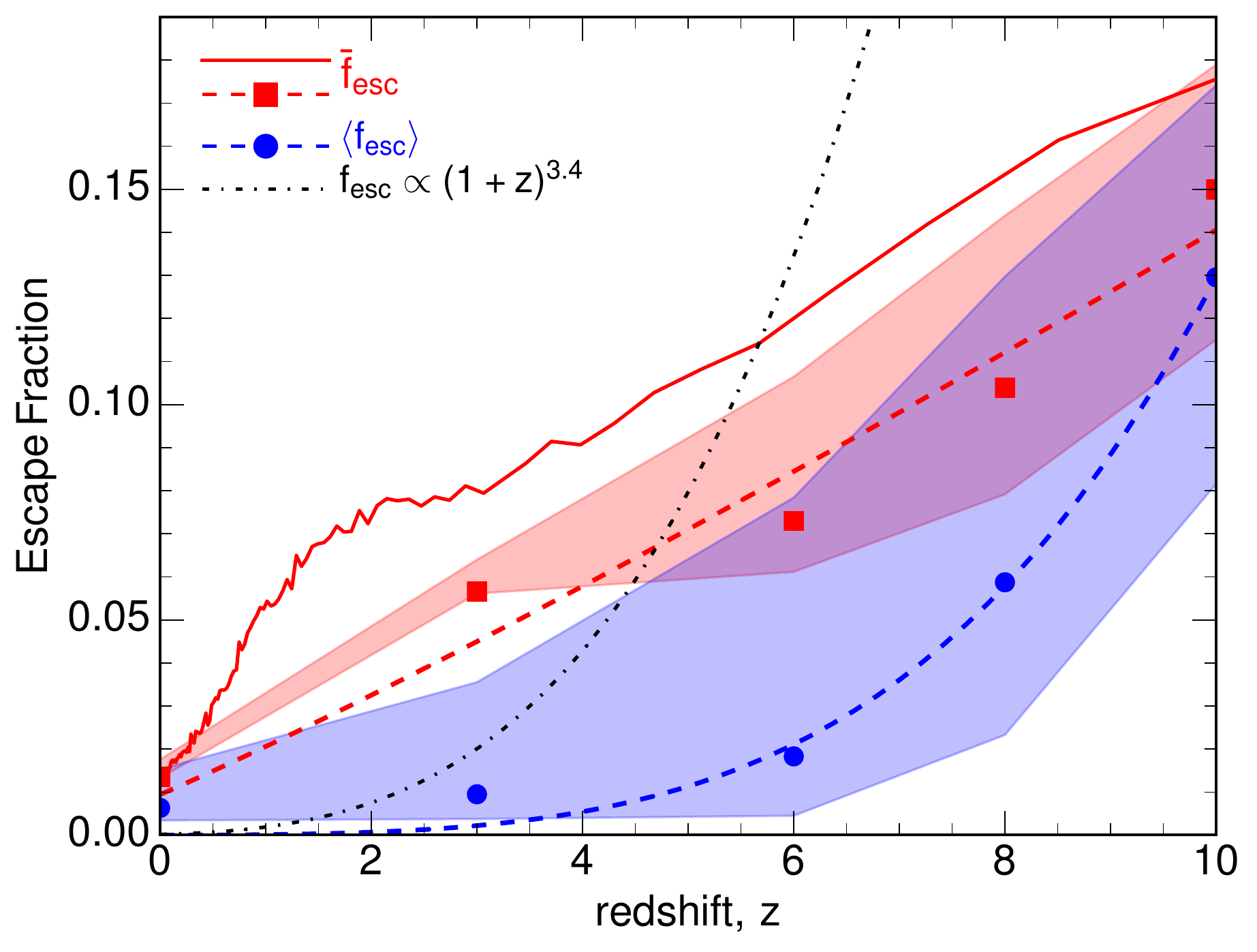}
 \caption{Evolution of the escape fraction of ionising photons for the
   population of \eagle\ galaxies. The luminosity-weighted mean escape
   fraction estimated from the model and the simulation, $\bar f_{\rm
     esc}$, is shown as a solid red curve. The evolution of the
   luminosity-weighted escape fraction computed by using the
   luminosity-dependent escape fraction at each redshift from the
   \eagle\ simulation, combined with the observed luminosity functions
   from \citet{Bouwens14} integrated down to a faint-end limit of
   $M_{\rm AB}=-13$ in 1500~\AA\ magnitude, is shown as red squares and fitted with a red dashed curve;
 the shaded regions show the range where the luminosity
   functions from \citet{Bouwens14} is extrapolated to either $-17$ to
   $-10$. The corresponding results for the galaxy number-weighted
   escape fraction, $\langle f_{\rm esc} \rangle$, are shown by the blue circles and shaded region; fitted with blue dashed curve.  The evolutionary trend from \citet{Haardt12} is shown
   as a black dash-dotted curve, and reaches $100$ percent at $z=
   10$. The escape fraction computed with our model evolves with
   redshift and achieves values $\gtrsim 10\%$ at redshifts, $z\gtrsim
   5$. The red curve determines the reionization and photo-ionisation
   history of the Universe, whereas the blue curve is the prediction
   for the escape fraction for typical observed galaxies at a given
   redshift.}
\label{fig_twofesc}
\end{figure}

In Fig.~\ref{fig_fesc_1} we plot the escape fraction as a function of the central surface density of star formation of a galaxy, $\dot\Sigma_{\star,0}\equiv
\dot M_\star/(2\pi R_\star^2)$, where $R_\star$ is the stellar
half-mass radius and $\dot M_\star$ the star formation rate of the
galaxy. For a disc in which $\sigmasfr$ falls exponentially with
radius, $\dot\Sigma_{\star,0}$ is approximately the central value of
the surface density of star formation. The escape fraction in the
simulations is well described by $f_{\rm esc}=0.2/(1+
\sigmasfrcrit/\csigmasfr)$, shown as a black dashed curve. There
is, however, a large amount of scatter at lower
redshifts.

Using this model, the escape fraction of \eagle\ galaxies is low or zero at low values
of $\dot\Sigma_{\star,0}\ll \sigmasfrcrit$, close to the maximum of
20~percent allowed by the model at high $\dot\Sigma_{\star,0}\gg
\sigmasfrcrit$, and exhibits a large scatter at intermediate values of
$\dot\Sigma_{\star,0}$. At $z=0$, however, a few rare examples do
reach the high values close to the maximum of 20~percent allowed by
our model. Since $\sigmasfr$ increases with $z$, galaxies in \eagle\
tend to have higher values of $f_{\rm esc}$ at increasing $z$. At
$z=3$, a considerable fraction has $f_{\rm esc}$ larger than a few
percent, and at $z=6$ most have $f_{\rm esc}>10$~percent.

The low values of $f_{\rm esc}$ at $z\approx 0$ are consistent with
those inferred observationally. For example, the Milky Way has an
escape fraction of $\approx 2$~percent \citep{Bland-Hawthorn01} and
\cite{Gnedin08} finds similarly low values for present day galaxies
(with many non-detections). The $z\approx 0$ galaxies discovered by
\cite{Borthakur14} and \cite{Izotov16,Izotov16b} with high escape fractions are
clearly exceptional, but such rare exceptions also occur in the
simulation where they correspond to vigorously star forming compact
galaxies - just as in the observational data.

Over the past decade or so considerable observational effort has been
made to detect ionising radiation emanating from LBGs at redshifts
$z\approx 3$ \citep[e.g.][]{Shapley06,Inoue06,Nestor13}, yielding
mostly non- or controversial detections. Similarly, most \eagle\
galaxies at $z=3$ have very low values of $f_{\rm esc}$
($<1$~percent), with fewer than 10~percent having $f_{\rm
  esc}>10$~percent that might be detectable observationally, in
reasonable agreement with the statistics presented by
\cite{Siana10,Nestor13}.

In Figure~\ref{fig_twofesc}, we plot the evolution of the UV
luminosity-weighted mean escape fraction, $\bar f_{\rm esc}$, of the
population of \eagle\ galaxies, computing the escape fraction for
individual star-forming regions as described above (solid red
line). The value of $\bar f_{\rm esc}$ increases from $\approx2$~percent at
$z=0$ to more than 10~percent above $z=5$, rising further towards
higher $z$. The trend is well described by $\bar f_{\rm esc}\propto(1+z)^{1.1}$ at $z>3$, with the increase in $\bar f_{\rm esc}$ with
$z$ much shallower than the $(1+z)^{3.4}$ assumed by \citet[black
dash-dotted curve in
Fig.~\ref{fig_twofesc}]{Haardt12}.
The figure also shows the evolution that results from combining the
dependence of $f_{\rm esc}$ on star formation rate from individual
galaxies from Fig.~\ref{fig_fesc_1} (black dashed line), with the observed
evolution of the luminosity function from \cite{Bouwens14},
extrapolated to $M_{\rm AB}=-13$ (where $M_{\rm AB}$ is the 1500\AA\
magnitude on the AB-system), either luminosity-weighted ($\bar f_{\rm
  esc}$, red dashed curve), or number-density weighted ($\langle
f_{\rm esc}\rangle$, blue curve); the shaded region corresponds to
extrapolating to $M_{\rm AB}=-17$ or $M_{\rm AB}=-10$. The luminosity-weighted mean escape fraction can be fitted by $\bar f_{\rm esc}=0.045~((1+z)/4)^{1.1}$ (red dashed curve) and the number-weighted mean by $\langle f_{\rm esc}\rangle=2.2\times10^{-3}~((1+z)/4)^{4}$ at $z>3$, and the maximum allowed value is 0.2. Since the escape fraction of galaxies in our model is much higher for vigorously star-forming galaxies, we have $\bar f_{\rm esc}> \langle f_{\rm
  esc}\rangle$. However, at higher $z$, most galaxies are highly star
forming and $\bar f_{\rm esc}\approx \langle f_{\rm
  esc}\rangle$. Interestingly, according to the blue curve, most of
the LBGs at redshift 3 are predicted to have low ($\lesssim 1$
per-cent) escape fractions.

Having demonstrated that our model yields the low values of escape
fractions observed directly at $z=0-3$, we proceed to investigate
whether $f_{\rm esc}$ at $z\gtrapprox 6$ is sufficient to reionize the
Universe, and whether the amplitude of the ionising background,
$\Gamma_{\rm HI}$, post-reionization is consistent with current
observations.

\section{Application to reionization}
\subsection{Emissivity and cumulative photon production}
\begin{figure}
 \centering
 \includegraphics[width=\linewidth]{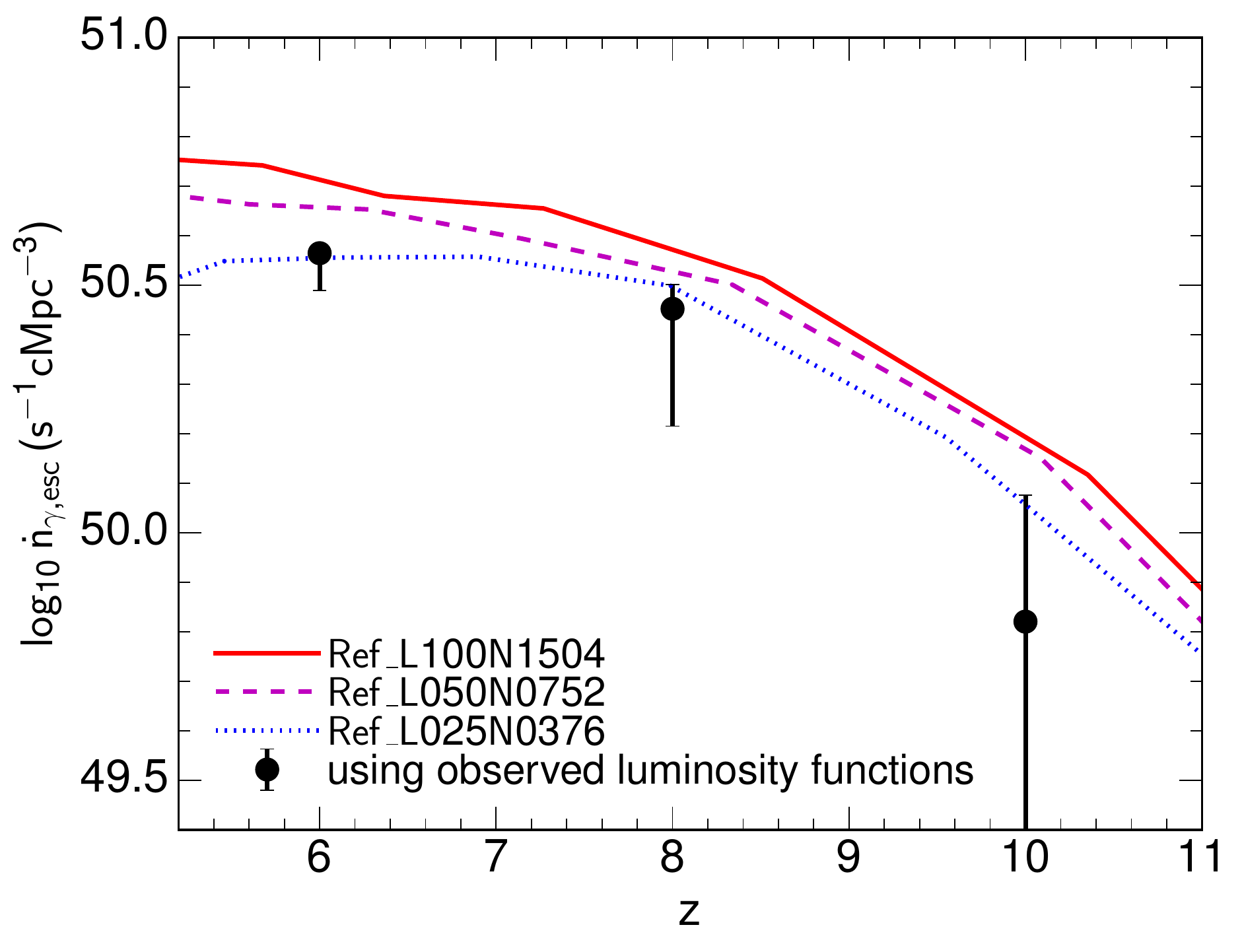}
 \caption{The rate at which ionising photons escape from galaxies
   per unit co-moving volume, $ \dot n_{\rm \gamma,esc}$, as a
   function of redshift using the \eagle\ galaxy stellar mass
   function (dotted blue, dashed magenta and solid red curves correspond to simulations
   Ref-L025N0376, Ref-L050N0752 and Ref-L100N1504,
   respectively). Black symbols combine the observed
   1500\AA~luminosity function extrapolated to $M_{1500, {\rm
       AB}}=-13$ from \citet{Bouwens14} with the evolution of the
   escape fraction taken from Fig.\ref{fig_fesc_1}; error bars span 
   the range covered if the faint end slope is extrapolated to $M_{1500, {\rm
       AB}}=-10$ and $M_{1500, {\rm
       AB}}=-16$.} \label{fig_emis}
\end{figure}

We use the population synthesis model of \cite{Schaerer03} to
calculate the total number of ionising photons produced by an
instantaneous starburst, per unit stellar mass formed, over the
lifetime of the stellar population, $dN_{\gamma}/dM_ \star$, as a
function of the initial metallicity. \cite{Schaerer03} assumes that
stars form with a Salpeter stellar initial mass function (IMF,
\citealt{Salpter55}) over the stellar mass range $1\hbox{--}100~{\rm
  M}_\odot$. However, in \eagle\ we assume that stars form with a
\cite{Chabrier03} IMF and over a stellar mass range $0.1\hbox{--}100$
M$_\odot$. We therefore divide Schaerer's value of
$dN_{\gamma}/dM_\star$ by a factor 2.55 to obtain
$dN_{\gamma}/dM_\star$ for stars forming in the range
$0.1\hbox{--}100~{\rm M}_\odot$ and then multiply again by a factor of
1.65 to convert from a Salpeter to a Chabrier IMF. This yields,
${dN_{\gamma}/ dM_\star} \approx 5\times 10^{60}\,{\rm M}_\odot^{-1}$,
consistent with the range $2\hbox{--}9\times 10^{60}$ M$_\odot^{-1}$
found in the study of \cite{Topping15}, which used a variety of
IMFs and included the effect of stellar rotation. This range is
probably a good estimate for the uncertainty in this value; the
contribution from binary stars is plausibly also important
\citep{Stanway16}.

The total number of s photons escaping galaxies, per unit
co-moving volume to redshift $z$, $n_{\rm \gamma, esc}$, is then
\begin{eqnarray}
	n_{\rm \gamma,esc}(z) &=&\int_0^\infty f_{\rm esc}(M_\star)\,{dN_{\gamma}\over dM_\star}\,\,M_\star\,n(M_{\star}, z)\,dM_\star\,
\end{eqnarray}
where $n(M_{\star}, z)$ is the comoving number density of galaxies
that formed a total stellar mass, $M_\star$, by redshift $z$.  From this
we can compute the emissivity, $\dot n_{\rm \gamma,esc}(z)\equiv
dn_{\rm \gamma,esc}/dt$ - the rate at which ionising photons escape
from galaxies per unit volume at redshift $z$. We can write this
emissivity also as $\dot n_{\rm \gamma, esc}\equiv \bar f_{\rm
  esc}\,\dot n_{\gamma,\star}$, in terms of the luminosity-weighted
mean escape fraction (see Fig.~\ref{fig_twofesc}) and the rate per
co-moving volume at which stars produce ionising photons, $\dot
n_{\gamma,\star}$.

We calculate $n(M_\star, z)$ from our three \eagle\ simulations and
plot $\dot n_{\rm \gamma,esc}$ as a function of redshift in
Fig.~\ref{fig_emis}. The emissivity is lower for the simulation of the
smallest volume (L025N0376) because it misses the more massive
galaxies that contribute significantly to $\dot n_\gamma$ in our
model. The emissivities for the two larger simulations are within
10~percent of each other, close enough that errors are dominated by
systematic uncertainties in computing $dN_\gamma/dM_\star$ and $f_{\rm
  esc}$ rather than $n(M_{\star}, z)$.

The ionising emissivities computed from the simulation also agree
reasonably well with those estimated from observations as $\dot n_{\rm
  \gamma,esc} = \bar f_{\rm esc}\ n_{\rm \gamma,\star,obs}$, with
$\bar f_{\rm esc}$ the fit to the escape fraction from \eagle\ found
in \S \ref{sec_model}, and $n_{\rm \gamma,\star,obs}$ the production
rate of ionising photons calculated by combining the integrated
1500\AA~luminosity functions as a function of redshift from
\cite{Bouwens14}, with the conversion factor from \cite{Schaerer03}
between 1500\AA~luminosity and ionising photon luminosity. This
agreement is not surprising since \eagle\ reproduces the observed
luminosity function relatively well, at least up to $z=6$
\citep{Furlong15a}.

\subsection{Timing of reionization}
\label{sec_Rz_2}
\begin{figure}
 \centering
 \includegraphics[width=\linewidth]{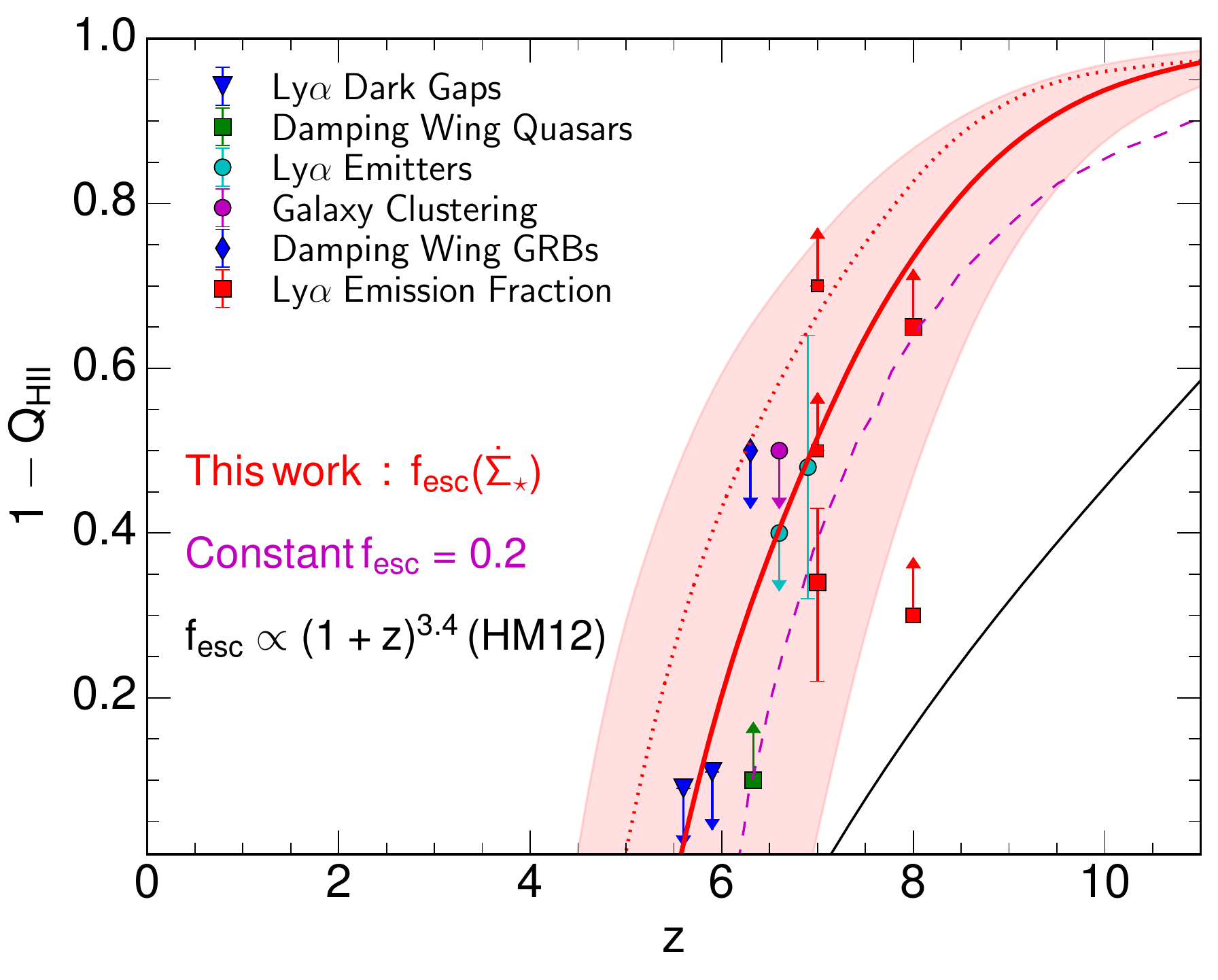}
 \caption{Evolution of the volume filling factor of ionized regions,
   $Q_{\rm HII}$, as a function of redshift. The solid red curve is
   for the \eagle\ simulation and uses our default values for the
   parameters of the model for the escape fraction, $f_{\rm esc,
     max}=20$~percent, $\sigmasfrcrit=0.1~{\rm M}_\odot~{\rm
     yr}^{-1}~{\rm kpc}^{-2}$. The red shaded region shows the effect
   of varying $f_{\rm esc, max}$ between 10 and 40~percent. The
   evolution of $Q_{\rm HII}$ found from combining the
   1500\AA~luminosity function from \citet{Bouwens14} with the escape
   fraction from our model is shown as the red dotted line; the model
   of \citet{Robertson15} which assumes $f_{\rm esc}=20$~percent for
   all galaxies is shown as the magenta dashed line; the model of
   \citet{Haardt12} is shown as the black solid line. These models are
   compared with observed estimates, using Ly$\alpha$ dark gaps
   statistics (blue triangles, \citealt{McGreer15}), the damping wing in
   a $z=7$ quasar (green square, \citealt{Mortlock11}), the damping wing in Gamma ray burst (black diamond, \citealt{Totani14}), galaxy clustering (magenta circle, \citealt{Mcquinn07}), Ly$\alpha$ emitters (cyan circles, \citealt{Ota08,Ouchi10}) and the
   Ly$\alpha$ emission statistics of galaxies
 {\protect\citep{Caruana12,Tilvi14,Schenker14}}.}
 \label{fig_Q}
\end{figure}

We calculate the evolution of the filling factor of ionised gas,
$Q_{\rm HII}$, by integrating Eq.~(\ref{eq:Q2}) using the results from
the \eagle\ simulation L0100N1504 for $\dot n_{\gamma, \star}$ and $\bar
f_{\rm esc}$ from Fig~\ref{fig_emis}, and the (extrapolated) evolution
of the clumping factor from \cite{Pawlik09}. The result is shown as
the red line in Fig.~\ref{fig_Q} for the default values of the
parameters of our model for the escape fraction, $f_{\rm esc,
  max}=20$~percent and $\sigmasfrcrit=0.1~{\rm M}_\odot~{\rm
  yr}^{-1}~{\rm kpc}^{-2}$. The ionised fraction is low at $z=9$,
when $Q_{\rm HII}\approx 10$~percent, reaches 50~percent by $z=7$, and
90~percent by $z=6$. This evolution is very similar to that obtained
from combining the observed 1500\AA~ luminosity function extrapolated
to $M_{1500, AB}=-13$ from \cite{Bouwens14} with our inferred evolution of
$\bar f_{\rm esc}$ (dotted red curve) and also to that of the model
from \cite{Robertson15} who take $f_{\rm esc}=20$~percent for all
galaxies (green curve). The transition from mostly neutral to mostly
ionised is much faster in all these models than in the model of
\cite{Haardt12} (solid black line). The red shaded region illustrates the
dependence of our model on the value of $f_{\rm esc, max}$, showing
the range obtained if the value is varied between 10 and 40~percent.

Inferred values for $Q_{\rm HII}(z)$ from observations are based on a
variety of methods which are all relatively indirect. Measurements
based on the statics of gaps with non-zero transmission in the
Ly$\alpha$ region of quasars \cite[e.g.][]{McGreer15}; on the damping
wing observed in a quasar spectrum \citep{Mortlock11} or a gamma-ray
burst spectrum \citep{Totani14}; and on the Ly$\alpha$ emission
properties of high-$z$ galaxies
\citep[e.g][]{Caruana12,Tilvi14,Schenker14} all suggest a relatively
rapid increase in $Q_{\rm HII}(z)$ from $z=8$ to $6$ (see also 
\citealt{Robertson15} and \citealt{Bouwens15}). These
values are uncertain, but they fit very well with the relatively rapid
evolution inferred from \eagle. \cite{Becker15} report the detection
of an extremely long and dark Ly$\alpha$ trough extending down to
$z=5.5$, which they argue is consistent with variations in the mean
free path expected to occur near the end of reionization. This
interpretation is consistent with the evolution of $Q_{\rm HII}(z)$
inferred from both the data and the \eagle\ model in Fig.~\ref{fig_Q}.

\subsection{Thomson optical depth}
\begin{figure}
 \centering
 \includegraphics[width=\linewidth]{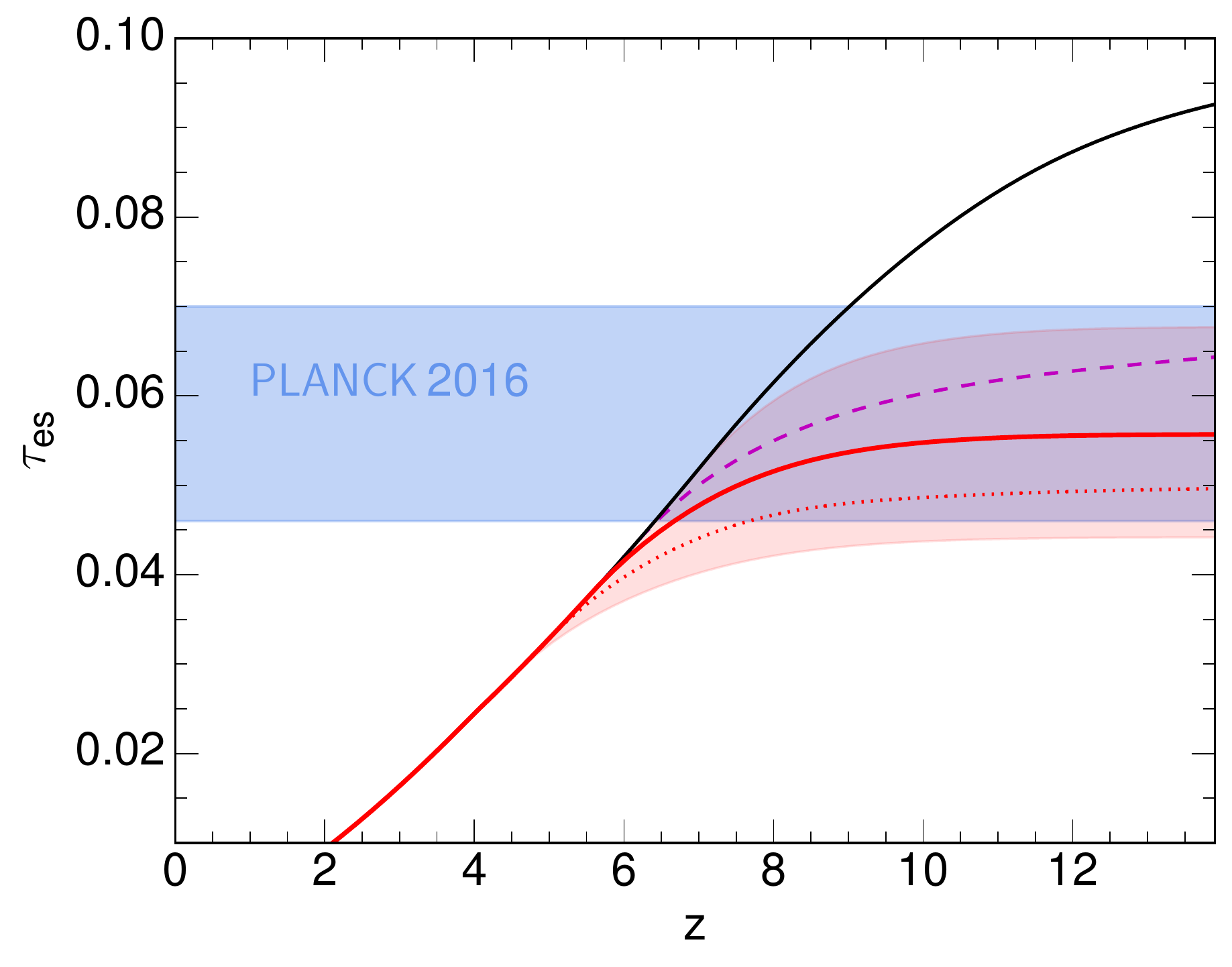}
 \caption{As Figure~\ref{fig_Q} but for the Thomson scattering optical
   depth, $\tau_{\rm es}$, as a function of redshift.  As before, the
   solid red line is for \eagle\ galaxies using our default parameters
   for the evolution of the escape fraction.  The horizontal shaded
   blue region demarcates the allowed $1~\sigma$ range from recent
   {\sc Planck} results (\citealt{Planck16}); the \eagle\ model is in
   good agreement with these data, as is the model of \citet{Robertson15},
   shown as the magenta dashed line, which assumes $f_{\rm
     esc}=20$~percent for all galaxies, By contrast, the model by
   \citet{Haardt12} (black solid line) overestimates $\tau_{\rm es}$.}
 \label{fig_tau}
\end{figure}

The optical depth due to Thomson scattering of cosmic microwave background photons off free
electrons, $\tau_{\rm es}(z)$, is a measure of the total column
density of free electrons between $z=0$ and a given redshift, and has
been measured from the Planck satellite data \citep{Planck16} for $z=z_{\rm
  CMB}\approx1100$, the redshift of the last scattering
surface. Within our model for reionization, it can be calculated from
the evolution of the ionised fraction $Q_{\rm HII}(z)$, as
\begin{equation}
	\tau_{\rm es}(z)=\int_{0}^{z}\,1.08~\sigma_{\rm T}~Q_{\rm HII}(z)~n_{\rm H}(z)\,~c~{H(z)}^{-1}~{\rm dz}\,,
\end{equation}
where $\sigma_{\rm T}$ is the Thomson cross section, and the factor
1.08 takes into account singly-ionized Helium. We plot $\tau_{\rm
  es}(z)$ for our default model of reionization with $f_{\rm esc,
  max}=20$~percent as the red curve in Figure~\ref{fig_tau}, as well
as a model based on the observed 1500\AA~ luminosity function
extrapolated to $M_{1500, AB}=-13$ from \cite{Bouwens14} with our
inferred evolution of $\bar f_{\rm esc}$ (solid red line) and the
model from \cite{Robertson15} which assumes $f_{\rm esc}=20$~percent
(magenta line). These all yield very similar results and all fall well
below the model of \cite{Haardt12} (solid black line). The models are
compared to the constraints from Planck \citep{Planck16} which apply
to $z=z_{\rm CMB}$ only, shown as the blue shaded region\footnote{The
  2016 Planck value of the optical depth is $\sim 20$~percent lower 
  than the 2015 value and is in much better agreement with our model
  predictions.}. The \eagle\ model is consistent with the {\sc Planck}
constraints. The effects of varying $f_{\rm esc, max}$ between 10 and
40~percent is illustrated by the red shaded region.

\subsection{The amplitude of the UV-background post-reionization}
\begin{figure}
 \centering
 \includegraphics[width=\linewidth]{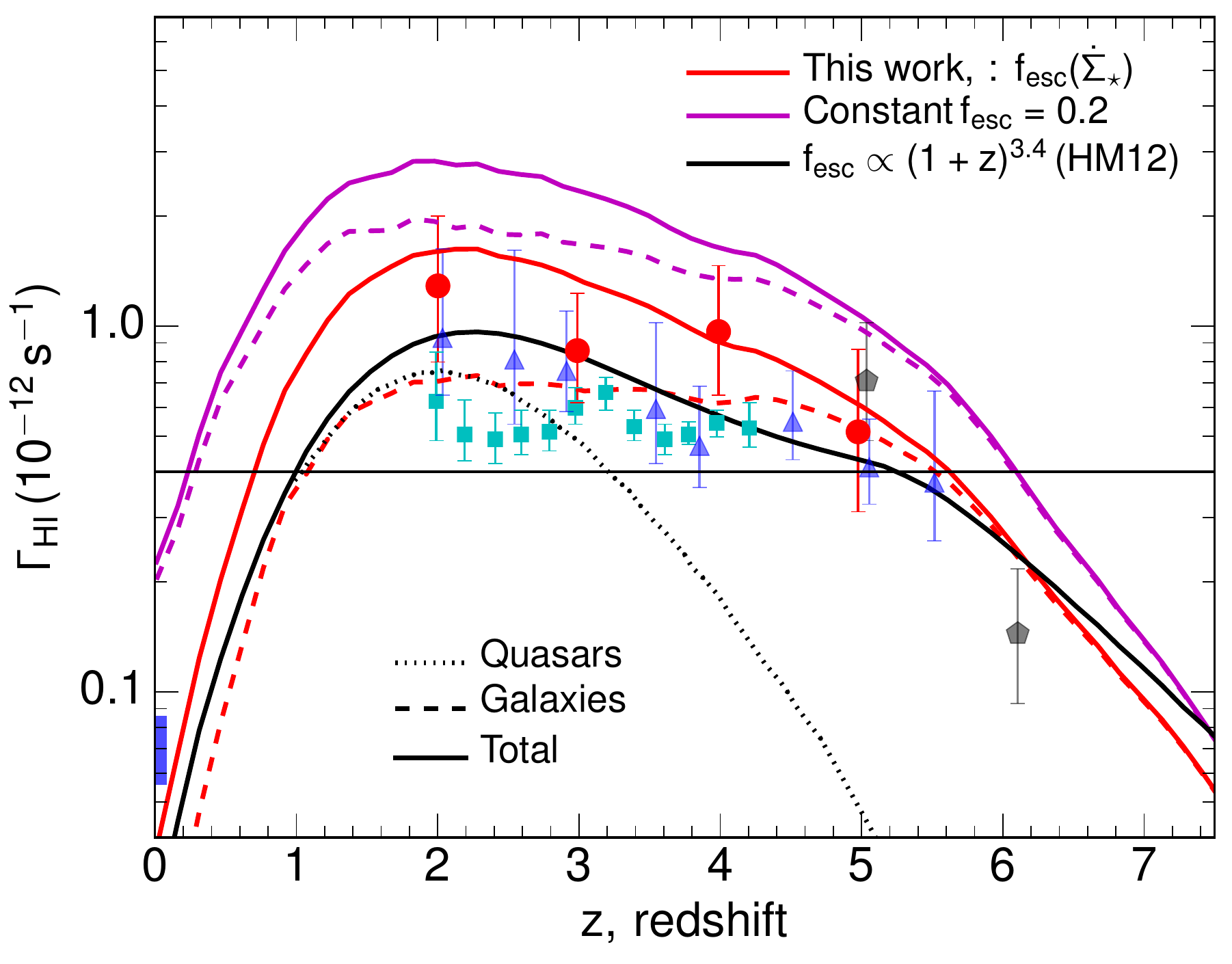}
 \caption{The {\rm HI} photo-ionisation rate, $\Gamma_{\rm HI}$, as a
   function of redshift. The dotted black curve is the contribution
   from quasars taken from \citet{Haardt12}, the dashed red line is
   the contribution from \eagle\ galaxies using the escape fraction
   from this paper, and the red line is the sum of both.  The
   photo-ionisation rate from \citet{Haardt12} is shown as the black
   line, and a model based on \eagle\ but using a constant value of
   $f_{\rm esc}=20$~percent is shown with magenta lines (dashed:
   galaxies only, full line: including QSOs).
 Various measurements based on the effective opacity in
   the Ly$\alpha$ forest are shown for comparison: red circles 
   \citep{Bolton07}; blue triangles \citep{Becker07} - cyan squares 
   \citep{Fau08} - measurements from the quasar near zones shown as grey circle
   \citep{Calverley11} -  measurement at $z\approx 0$ shown as blue vertical bar \citep{Fumagalli17}.}
 \label{fig_J21}
\end{figure}

Models for the evolution of the escape fraction should also reproduce
the observed photo-ionisation rate, $\Gamma_{\rm HI}$, after
reionization. The photo-ionisation rate is inferred observationally
from measuring the mean transmission in the Ly$\alpha$ forest of
high-$z$ QSOs, but that quantity is degenerate with the poorly
constrained temperature of the intergalactic medium and the level of
small-scale clustering of gas \cite[e.g.][]{Rauch97}. The
photo-ionisation rate is related to the ionising emissivity through the
mean free path of ionising photons, which is closely related to the mean
distance between Lyman limit systems. We use the method that relates
emissivity and photo-ionisation rate developed by \cite{Haardt12} to compute
$\Gamma_{\rm HI}$ from \eagle\ galaxies. Adding the contribution from
QSOs computed by those authors, we compare the net rate to other model
predictions as well as observations in Figure~\ref{fig_J21}.

The value of $\Gamma_{\rm HI}$ obtained by combining the contribution
of \eagle\ galaxies with that of QSOs rises by a factor of 10 between
$z=8$ and $z=4$, remains constant to within a factor of 2 to $z=1.5$,
and then drops rapidly towards $z=0$. The contribution of galaxies
dominates above $z=2$, is close to 90~percent of the total at $z=4$
and increases even further towards higher $z$. Our values are higher than
those of \cite{Haardt12} because of the shallower evolution of the
escape fraction in our case. Our calculations of $f_{\rm esc}$ do not
consider absorption by dust, which should play an increasingly
important role at lower $z$. We may therefore increasingly
overestimate $\Gamma_{\rm HI}$ towards lower $z$.

\cite{Bolton07} and \cite{Becker07} quote uncertainties in the
measured values of $\Gamma_{\rm HI}$ of 50~percent or more but, in
addition, there appear to be systematic differences in their values
compared to those of \cite{Fau08}. The values we infer from \eagle\ at
$z=2-3$ agree very well with the data of \cite{Bolton07}, are slightly
higher than the data of \cite{Becker07} and, at best, are marginally
consistent with the data of \cite{Fau08} which have the smallest
error bars. The agreement with these datasets improves at higher $z$,
but our model predictions are somewhat above the value inferred by
\cite{Calverley11} from the quasar near zone measurement at
$z\approx 6$.
 	
A model in which $f_{\rm esc}=20$ percent (magenta line), similar to
that discussed by \cite{Robertson15} and \cite{Bouwens15} yields
similar values of $\Gamma_{\rm HI}$ to those from our \eagle\ model
(red curve) down to $z\approx 7$, but yields values that become
increasingly high at lower redshifts. At $z=0$ such a model predicts
$\Gamma_{\rm HI}(z=0)\approx 2\times 10^{-13}~{\rm s}^{-1}$, more than
an order of magnitude above the $z=0$ upper limit of
\cite{Adams11}. This demonstrates once again that the escape fraction
{\em has} to vary with $z$, in order to reconcile the low values of
$\Gamma_{\rm HI}$ measured below $z=1$, say, with the relatively high
values inferred above $z=5$ (see also \citealt{Khaire15}). However the
evolution cannot be as steep as that proposed by \cite{Haardt12},
$f_{\rm esc}\propto (1+z)^{3.4}$, because that conflicts with the
evolution of $Q_{\rm HII}$ and the measured $\tau_{\rm es}(z=z_{\rm
  CMB})$ discussed above.

The photo-ionisation rate due to quasars alone from \citet{Haardt01}, shown as a dotted line in Fig.~\ref{fig_J21}, decreases rapidly with increasing redshift above $z\sim 2$ due to the steep decline in the number density of (relatively) bright quasars. In a recent study \cite{Giallongo15} suggest that the number density of faint quasars decreases much less rapidly beyond $z=3$. Adding the contribution of such faint quasars, using the fit proposed by \cite{Giallongo15}, has a relatively small effect, less than 10~percent on the value of $\Gamma_{\rm HI}$ in our fiducial model (red solid line in Fig.~\ref{fig_J21}). \cite{Haardt15} extrapolate the data presented by \cite{Giallongo15} and claim that such faint quasars by themselves emit enough ionising photons to ionise the Universe (see also \citealt{Mitra16}). However, adding the contribution that results from this extrapolation, taken from Fig.~3 in \cite{Haardt15}, would double the value of $\Gamma_{\rm HI}$ at $z=5$, and almost triple it at $z=6$. Such high values may be in tension with observations, especially at $z=6$. Future observations
	should clarify the contribution of quasars to $\Gamma_{\rm HI}$ and to reionization.

\section{Summary}
\label{sec_Dis}
The fraction of ionising photons that escapes from galaxies is not
well constrained observationally and theoretical predictions vary
widely. The low values of the escape fraction measured locally and
the strong indication that the Universe reionized close to redshifts
$z=6-8$ requires $f_{\rm esc}$ to evolve strongly with $z$,
increasing from a few per cent at $z=0$, to $\sim 20$~per cent above
$z=6$. Here we presented a model based on the \eagle\ simulations that
explains this evolution. 

In the \eagle\ simulations, the median density at which stars form increases rapidly
with redshift. This implies that stars form at increasingly
higher pressure, or equivalently, star formation surface density,
$\sigmasfr$, and yields smaller galaxies with higher specific
star formation rates with increasing $z$. Physically this is a
consequence of the larger cosmological accretion rate onto galaxies
at higher $z$, combined with the evolution of the efficiency of
feedback from star formation. Such a trend is also seen
observationally since galaxy specific star formation rates increase
rapidly with $z$ whereas galaxy sizes decrease. Indeed in
\cite{Sharma15} we showed that \eagle\ reproduces the observed
evolution of star formation surface densities.

Star formation at high values of $\sigmasfr$ is observed to result in
strong galactic winds \citep[e.g.][]{Heckman11}, and indeed, such
winds, although rare at $z=0$, are ubiquitous at higher $z$
\citep[e.g.][]{Pettini02,Weiner09,Bradshaw13}. Following
\cite{Sharma15}, we assumed that winds increase the fraction of
ionising photons that can escape their galaxy through channels carved
by outflows through the neutral gas in the interstellar medium. Our
model has two main parameters: $\sigmasfrcrit$, the star formation
rate surface density above which strong winds are launched, and
$f_{\rm esc, max}$, the escape fraction in the presence of such
winds. Therefore, the escape fraction is large in compact starbursts;
there is some observational evidence supporting this
\citep{Borthakur14, deBarros15, Izotov16,Izotov16b}. As in \cite{Sharma15} we use
$\sigmasfrcrit=0.1~{\rm M}_\odot~{\rm yr}^{-1}~{\rm kpc}^{-2}$, as
suggested by \cite{Heckman01} and $f_{\rm esc, max}=20$~per cent,
motivated by the observations of \cite{Borthakur14} of the escape
fraction in a $z\sim 0$ compact starburst.

Applying this model to star formation in \eagle\, we find that $f_{\rm
  esc}$ increases rapidly with redshift, from very low values at $z=0$
to nearly 20~per cent at $z>6$. This occurs because very few galaxies
form stars with $\sigmasfr\ge \sigmasfrcrit$ at low $z$, whereas most
do so at $z>6$ (Fig.~\ref{fig_sigsize}). As a consequence, $f_{\rm
  esc}$ evolves rapidly enough to yield a realistic evolution for the
filling factor of ionized gas (Fig.~\ref{fig_Q}), the Thomson optical
depth to the surface of last scattering (Fig.~\ref{fig_tau}), and the
amplitude of the ionising background post-reionization
(Fig.~\ref{fig_J21}).  In particular, we find that:

\begin{enumerate}
\item the escape fraction depends strongly on the central surface
  density of star formation, $\csigmasfr$, approximately as $f_{\rm
    esc}=0.2/(1+\sigmasfrcrit/\csigmasfr)$ (Fig.~\ref{fig_fesc_1}).
\item the luminosity-weighted mean escape fraction in the absence of
  dust, averaged over the galaxy population,  evolves as $\bar f_{\rm esc}=0.045~((1+z)/4)^{1.1}$ and becomes constant at $\approx0.2$ at redshift $z>10$. The galaxy number-weighted mean escape fraction of galaxies as a function of redshift evolves as $f_{\rm
  esc}=2.2\times10^{-3}~((1+z)/4)^{4}$ (Fig.~\ref{fig_twofesc}).
\item the escape fraction is, in general, higher for brighter
  star-forming galaxies. As a consequence faint galaxies below the HST UDF detection
  limit do not dominate the photon budget for reionization
  \citep{Sharma15}.
\end{enumerate}

A complete theoretical understanding of the physics governing the
dependence of the escape fraction on $\sigmasfr$ is lacking. It would
require accurate radiation-hydrodynamic modelling of the interstellar
medium and of individual star forming regions, including the radiative
and energetic feedback from massive stars and supernovae. Recent
studies have achieved some progress towards this goal by concentrating
on small galaxies at high $z$ \citep[e.g.][] {Wise09,Kimm14,Ma15},
although even these lack the resolution and physics required to
capture the intricacies that influence the escape fraction at the
scale of molecular clouds.

In the case where massive stars carve the channels through which
photons can escape, there may still be a timing issue
preventing a galaxy from having a high emissivity. Indeed, the
production of ionising photons may have decreased
significantly already before the channels are opened, given the short
lifetime of massive stars. The emissivity may then be low when the
escape fraction is high. Ionising photons emitted by binary stars
\citep{Stanway16} may help overcome this \citep[see also][]{Ma16}.

We have neglected the effects of dust, except in the calculation of
$\Gamma_{\rm HI}$. Dust likely has a negligible effect at high $z$
($z>6$, say), given that observed sources at high redshift are dust
poor \citep{Bouwens14}. Dust may, however, play a much more important
role at lower redshifts, potentially explaining the fact that many
low-redshift detections report low values of $f_{\rm esc}$ despite the
highly eruptive nature of many of the galaxies investigated
\citep{Leitet13}. The escape fraction reported by \citet{Borthakur14}
is reduced from $20$ to $\approx2$ percent if the effect of dust is
taken into account.

Reionization in our model proceeds quite rapidly, much faster than in
the model of \cite{Haardt12}, as illustrated in Fig.~\ref{fig_Q}. One
reason for this is that the escape fraction in the simulations
declines more slowly with decreasing redshift than in the
\cite{Haardt12} model, giving rise to a steeper build up of the
emissivity. The rate of evolution of the emissivity (and thus the
speed at which reionization occurs) depends, of course, on the rate of
evolution of the galaxy luminosity function.  Both in our simulations
and in the data, the bright end of the luminosity function evolves
faster than the faint end \citep[e.g][]{Bouwens14}. Since the
production of ionising photons in our model is dominated by brighter
galaxies \citep{Sharma15}, the emissivity and the volume filling
factor, $Q_{\rm HII}$, evolve quite rapidly.  

{Here we find that the brighter galaxies dominate reionization (see also \citet{Sharma15}), in apparent contradiction with studies that claim that {\em faint} galaxies are the drivers of reionization \citep[e.g.][]{Yajima11, Wise14}. In the model by
	\cite{Wise14}, galaxies that contribute to reionization have $M_{1500}\sim -17$ (their Fig.~15) and are hosted by halos of mass
	$M_h\approx 10^8{\rm M}_\odot$ (their Fig.~8) at $z=7$. These values, in fact, agree well with our findings (see Fig.~\ref{fig_fesc_mult}). We also agree that lower mass galaxies
	with lower star formation rates contribute progressively more with increasing $z$. The reason for the difference in interpretation,
	notwithstanding the similarity in results, is twofold: ({\em i}) Reionization is relatively rapid in our model, with the volume filling factor of ionised gas increasing from $Q_{\rm HII}=0.2$ at $z=8$ to $Q_{\rm HII}=0.8$ at $z=6$. In fact, half of all ionisations occurring below $z=7$,  galaxies that
	dominate the ionising emissivity at $z\gtrapprox 8$ are therefore not particularly relevant for {\em reionization} itself. ({ii}) We
	call the $z\sim 7$ galaxies brighter than $M_{1500}=-17$ bright because they are comfortably above the HST detection limit. It is
	in this sense that we claim that the brighter galaxies reionized the Universe: our model does not need to appeal to a putative
	population of galaxies much fainter than current (or future) detection limits, for  galaxies to be able to reionise the Universe.}

Existing observational data, displayed in Fig.~\ref{fig_Q}, support
the steep history of reionization as obtained in this work.  (Radiative transfer
calculations suggest that reionization might have slowed down near the
end due to the effect of LLSs, e.g. \citealt{Shukla16}). Future observational surveys should constrain the evolution of
the ionised fraction better and thus shed light on the nature of the sources
that reionized the Universe.

An {\em ab initio} calculation of the escape fraction remains an
important, if challenging, goal of theoretical studies of
reionization. Our model, in which the escape of photons is determined by the ability of galaxies to drive winds, exhibits the right
phenomenology to explain why $f_{\rm esc}$ evolves rapidly with $z$
and may help guide such studies. From an observational perspective,
the inference of $f_{\rm esc}$ from the detection of nebular emission lines in early star forming galaxies
with JWST \cite[e.g.][]{Erb16} will provide an important test of our model.

\section*{Acknowledgments}
We are grateful to Lydia Heck and  Peter Draper for supporting with their expertise in high performance computing.  We thank PRACE for the access to the Curie facility in France. We have used the DiRAC system which is a part of National E-Infrastructure at Durham University, operated by the Institute for Computational Cosmology on behalf of the STFC DiRAC HPC Facility (www.dirac.ac.uk); the equipment was funded by BIS National E-infrastructure capital grant ST/K00042X/1, STFC capital grant ST/H008519/1, STFC DiRAC Operations grant ST/K003267/1 and Durham University. 
The study was sponsored by the Dutch National Computing Facilities Foundation (NCF) for the use of supercomputer facilities, with financial support from the Netherlands Organisation for Scientific Research (NWO), and the European Research Council under the European Unions Seventh Framework Programme (FP7/2007- 2013) / ERC Grant agreements 278594 GasAroundGalaxies, GA 267291 Cosmiway, and 321334 dustygal. Support was also received via the Interuniversity Attraction Poles Programme initiated by the Belgian Science Policy Office ([AP P7/08 CHARM]), the National Science Foundation under Grant No. NSF PHY11-25915, and the UK Science and Technology Facilities Council (grant numbers ST/F001166/1 and ST/I000976/1) via rolling and consolidating grants awarded to the ICC. RAC is a Royal Society university research fellow. M.Sharma is an STFC Post-doctoral fellow at the ICC. Some of the data in this paper is available in the \eagle\ database \cite{McAlpine16} or through the authors.



\footnotesize{\bibliography{ref_eagle_He}}
\end{document}